\documentclass[letterpaper,twocolumn,10pt]{article}

\usepackage{mdframed}
\usepackage{adjustbox}
\usepackage{multirow}
\usepackage[table, dvipsnames]{xcolor}
\usepackage{usenix-2020-09}
\usepackage{dtrt}
\usepackage{tikz}
\usepackage{stmaryrd} 
\usepackage{authblk}
\usepackage{subcaption}
\usepackage{lineno}
\usepackage{mathtools}
\usepackage{amsmath}
\usepackage{amssymb}
\usepackage{inconsolata}
\usepackage{graphicx}
\usepackage{amsfonts}
\usepackage{booktabs}
\usepackage[flushleft]{threeparttable}
\usepackage{authblk}
\usepackage{pifont}
\usepackage{pgfplots}
\usepackage[stable]{footmisc}
\usepackage{makecell}
\usepackage[utf8]{inputenc}
\usepackage{listings}
\usepackage{semantic}
\usepackage{listings-coq}
\usepackage{listings-rust}
\usepackage{listings-sql}
\usepackage{xspace}
\usepackage{amsthm}
\usepackage[russian,english]{babel}
\usetikzlibrary{fit, positioning, arrows.meta}
\newcommand{\picachv}{\textsc{Picachv}\xspace}
\theoremstyle{definition}
\newtheorem{definition}{Definition}[section]
\newtheorem{theorem}{Theorem}[section]

\definecolor{dark-red}{rgb}{0.4, 0.15, 0.15}
\definecolor{dark-blue}{rgb}{0.15, 0.15, 0.4}
\definecolor{medium-red}{rgb}{0.5, 0, 0}
\definecolor{medium-blue}{rgb}{0, 0, 0.5}
\definecolor{light-red}{rgb}{0.7, 0, 0}
\definecolor{light-blue}{rgb}{0, 0, 0.7}

\hypersetup{
  colorlinks,
  linkcolor={dark-red},
  citecolor={dark-blue},
  urlcolor={medium-blue}
}
\begin{document}
\date{}


\title{\picachv: Formally Verified Data Use Policy Enforcement for Secure Data Analytics}


\author[1]{Haobin Hiroki Chen}
\author[1]{Hongbo Chen}
\author[2]{Mingshen Sun}
\author[1]{Chenghong Wang}
\author[1]{XiaoFeng Wang}
\affil[1]{\it Indiana University Bloomington}
\affil[2]{\it Independent Researcher}

\maketitle
\renewcommand{\thefootnote}{\arabic{footnote}}
\thispagestyle{plain}
\begin{abstract}
Ensuring the proper use of sensitive data in analytics under complex privacy policies is an increasingly critical challenge. Many existing approaches lack portability, verifiability, and scalability across diverse data processing frameworks. We introduce \textsc{Picachv}, a novel security monitor that automatically enforces data use policies. It works on relational algebra as an abstraction for program semantics, enabling policy enforcement on query plans generated by programs during execution. This approach simplifies analysis across diverse analytical operations and supports various front-end query languages. By formalizing both data use policies and relational algebra semantics in Coq, we prove that \textsc{Picachv} correctly enforces policies. {\textsc{Picachv}} also leverages Trusted Execution Environments (TEEs) to enhance trust in runtime, providing provable policy compliance to stakeholders that the analytical tasks comply with their data use policies. We integrated \picachv into \texttt{Polars}, a state-of-the-art data analytics framework, and evaluate its performance using the TPC-H benchmark. We also apply our approach to real-world use cases. Our work demonstrates the practical application of formal methods in securing data analytics, addressing key challenges. 

\end{abstract}
\newenvironment{proofsketch}{%
  \renewcommand{\proofname}{Proof Sketch}\proof}{\endproof}
\renewcommand{\subfigureautorefname}{\figureautorefname}
\newcommand{\definitionautorefname}{Definition}%

\renewcommand{\sectionautorefname}{Section}%
\renewcommand{\subsectionautorefname}{Section}%
\renewcommand{\subsubsectionautorefname}{Section}%
\newcommand{\hongbo}[1]{\dtcolornote[HC]{blue}{#1}}
\newcommand{\hiroki}[1]{\dtcolornote[Hiroki]{red}{#1}}
\newcommand{\cw}[1]{\dtcolornote[Chenghong]{brown}{#1}}

\newcommand*\circled[1]{\tikz[baseline=(char.base)]{
            \node[shape=circle,draw,inner sep=2pt] (char) {#1};}}
\newcommand\denote[1]{\left\llbracket #1 \right\rrbracket}
\newcommand\tuple[1]{\left\langle\ #1 \ \right\rangle}
\newcommand\step[1]{\lhook\joinrel\xrightarrow{#1}}
\newcommand\noindentsec[1]{\vspace{.25em}\noindent\textbf{#1}}
\newcommand\flowsto[2]{#1 \sqsubseteq^{\ast} #2}
\newcommand\coloredbg[2]{\text{\colorbox{#1 !30}{$#2$}}}

\section{Introduction}
\label{sec:intro}
The rapid advancement of computational devices and the rise of big data present unparalleled opportunities to accelerate scientific progress and drive innovation through data-driven decision-making. While these opportunities are ground-breaking, they are accompanied by a critical challenge: ensuring proper data \textit{use} in compliance with complex privacy policies. Real-world data analytics often involves complex privacy policies. For instance, researchers using NIH’s \textit{All of Us}~\cite{allofus} platform must comply with multiple privacy regulations, such as the HIPAA Safe Harbor Rule~\cite{hipaa} and the platform’s own policies mandating patient data aggregation into groups of 20 or more.

Unfortunately, enforcing these policies poses a great challenge. Manual checks are impractical due to the complexity of analytical tasks and the high cost of human effort. This leads to reliance on machine-based solutions, which often fail to address these challenges comprehensively. For example, database access control mechanisms~\cite{qapla, ifdb, blockaid, multiverse-db}, focus solely on restricting access to data but do not ensure that \textit{already authorized personnel} comply with proper data use policies. Moreover, they lack verifiable guarantees to data holders as they frequently rely on very \textit{ad hoc} heuristics. Program analysis~\cite{jacqueline, jeeves, estrela, privguard, selinq, storm, urflow, daisy} mostly still focus on access control policies. The few that can be or have been extended to data use policy enforcement often restrict programmers to a specific language (e.g., \textsc{PrivGuard}~\cite{privguard} only supports Python), which is less desirable.

Addressing this problem remains non-trivial. One major challenge is the diversity of front-end programming languages. In data analytics, researchers and data scientists use a wide range of languages and rely heavily on third-party libraries. Some frameworks like Apache Spark~\cite{spark} supports both Dataframe-like APIs and SQL. 
This diversity in frontend programming languages and tools creates significant challenges for implementing consistent and universal policy enforcement mechanisms. Each language has its own syntax, semantics, and idiosyncrasies. This variability complicates the process of policy enforcement, as a mechanism tailored for one language may not be applicable to others. Moreover, complex third-party libraries often obscure underlying data operations, making policy enforcement more challenging. Traditional approaches that focus on language-specific static or dynamic analysis techniques \cite{storm, jacqueline, urflow, daisy, jeeves} struggle to provide a comprehensive solution across this diverse landscape.

Enforcing data use policies essentially relies on \textit{information flow control} (IFC)~\cite{ifc-sok, ifdb}, with the underexplored aspect of \textit{declassification} playing a key role. Existing works on declassification~\cite{downgrading, downgrade-non-inter, declassification} primarily focus on language-level approaches and provide only general methodologies for specifying policies, but they do not address the specific requirements of our use case. Here, several key challenges emerge: 1) accurately describing data sensitivity levels and downgrading operations for data analytics,  2) proposing new semantics for the relational algebra that interplays with the declassification requirements, and then 3) defining and proving a rigorous \textit{semantics-based security model}~\cite{language-based-ifc} for policy enforcement. Addressing these design complexities requires formal verification to provide mathematically rigorous security guarantees, which are critical for highly sensitive workloads.

Beyond security guarantees, effective policy enforcement must also account for performance trade-offs and system scalability. For analytical queries, performance (or total cost of ownership (TCO)) is always a critical concern in policy enforcement, with both static and dynamic approaches facing unique challenges. While static analysis minimizes runtime overhead, it introduces significant development costs~\cite{knowledge-based-ifc} and lacks the precision (i.e., high false positive rate) achievable with dynamic analysis. Dynamic analysis, on the other hand, grapples with the significant runtime complexity of tracking tags across data, and the tags quickly become cumbersome and unmanageable as the system scales up. These performance considerations necessitate careful balancing of enforcement strategies.

\vspace{0.25em}\noindent\textbf{This work.}
We tackle these challenges from a different perspective: relational algebra. Data analysis programs, regardless of their implementation language or framework, can be logically described using relational algebra or similar intermediate representations. Such translations have already been well-studied~\cite{pytond, snakes, python-sql}, there are even extensions for other applications like ML inferences~\cite{ra-exntension}. Hence, we assume the existence of such techniques and focus our research on designing enforcement mechanisms over relational algebra (often in the form of query plans). We therefore describe all data manipulations in a unified query plan, overcoming the limitations imposed by specific languages or frameworks as we obtain the semantics of the program via relational algebra. Furthermore, this insight aligns seamlessly with current policy requirements, which specify necessary operations before release, as these policies inherently describe \textit{data manipulation requirements} that can be integrated into relational algebra.

While policy enforcement at this level seems straightforward, there is a question left unsolved: How should we structure the security lattice and declassification rules to express diverse data use policies? Existing works on declassification policies~\cite{knowledge-based-ifc, downgrade-non-inter, downgrading} provide only high-level methodologies and are not directly applicable to our scenarios.
Fortunately, data use policies inherently imply sensitivity ordering through manipulation requirements. For example, privacy policies like redaction or aggregation naturally reduce data sensitivity to some extent, giving us an intuition for structuring the security lattice in alignment with these data operations. This work presents a feasible way for specifying data use policies based on this key observation. On top of that, we propose in this paper a new operational semantics for relational algebra to support policy enforcement. To ensure correctness, we rigorously verify these semantics using Coq, providing mathematically sound guarantees for policy compliance.

We implement a prototype runtime security monitor, called \picachv. Informally, \picachv functions as a verified middleware that intercepts query plans and monitors query execution in parallel to detect policy violations. Since \picachv focuses on the relational data, it allows for sophisticated optimization techniques such as parallelization. Furthermore, \picachv integrates with TEEs to provide strong \textit{proofs} to data owners that data is used appropriately during analytical tasks in untrusted cloud environments. TEEs provide cryptographic reports via remote attestation, ensuring the authenticity of the monitor, the integrity of operations, and proving policy compliance of queries to the data owner. Nevertheless, our technique can be applied to trusted environments like on-premise databases.

We summarize our contributions as follows. 
\begin{itemize}\setlength\itemsep{0em}
    \item We formalize relational operators with declassification semantics to avoid the complexity of language-based semantics. We formalize the semantics and also provide the mechanized proof of soundness in Coq.
    \item We design \picachv, a dynamic security monitor that transparently enforces these data use policies when executing analytical tasks. Our framework also incorporates multiple optimizations, and experiments show that only small runtime overhead was incurred.
    \item We ported \picachv to a well-established data analysis framework called \texttt{Polars} and performed evaluation to showcase the adaptability of \picachv in diverse data processing environments.
\end{itemize}

\noindentsec{Roadmap.} This paper is structured as follows: In \autoref{sec:related}, we review related work. \autoref{sec:preliminary} introduces essential background and preliminary knowledge to provide readers with the necessary foundation. In \autoref{sec:formal}, we formalize the data use policies central to our approach. \autoref{sec:ra} details the operational semantics of our policy-integrated relational algebra. \autoref{sec:implementation} outlines the implementation of \picachv. \autoref{sec:evaluation} evaluates \picachv, demonstrating its effectiveness and efficiency across various scenarios. Finally, \autoref{sec:discussion} discusses the scalability, performance trade-offs, and potential limitations of our approach.

\section{Related Work}
\label{sec:related}
\vspace{0.25em}\noindent\textbf{Policy enforcement in data analytics.} Several frameworks have been developed to analyze application code that interacts with sensitive data. Language-based information flow control mechanisms, such as \textsc{Storm}\cite{storm}, \textsc{RuleKeeper}\cite{rulekeeper}, \textsc{UrFlow}\cite{urflow}, \textsc{SeLINQ}\cite{selinq}, \textsc{Swift}\cite{swift}, \textsc{PrivGuard}~\cite{privguard} and \textsc{Jeeves}\cite{jeeves} renforce policies by rejecting non-compliant code at compile time. Some solutions, however, require manual code annotations, which can be labor-intensive and error-prone. Moreover, these frameworks struggle to enforce fine-grained policies effectively compared to  \textsc{Picachv}. Query rewriting approaches, including \textsc{Qapla}\cite{qapla}, \textsc{Jacqueline}\cite{jacqueline}, and \textsc{Estrela}~\cite{estrela}, aim to ensure policy compliance by modifying query syntax. However, these methods operate at the syntax level, often sacrificing the precision required for execution-time analysis. The most similar work to \picachv is \textsc{Laputa}~\cite{laputa}, which enforces table-level policies for Apache SparkQL using regular expression matching on physical query plans. However, like many earlier efforts, \textsc{Laputa} relies heavily on heuristics and lacks verifiable guarantees, limiting its applicability for sensitive workloads. It also lacks fine-grained policy enforcement.

\vspace{0.25em}\noindent\textbf{Formal methods.} Formal methods are essential for ensuring the correctness of critical systems by providing mathematically rigorous guarantees and eliminating reliance on \textit{ad hoc} testing. Large-scale systems such as operating system kernels and virtual machine managers have successfully adopted formal verification techniques~\cite{verified-kvm, verified-browser, verified-hypervisor}, demonstrating their industrial feasibility. Applying formal methods in securing data analytics remains a rare practice. Existing research has formalized and verified the correctness of database systems~\cite{verified-db} and SQL semantics~\cite{sql-semantics}, but no work has yet addressed the integration of privacy policies into relational algebra.

\vspace{0.25em}\noindent\textbf{TEEs for secure data computation.}
Confidential Computing enables secure data processing, unlocking new computing scenarios and fostering innovation across industries where privacy concerns previously posed barriers. Under the hood, TEEs protect unauthorized parties from seeing and modifying the data and code inside TEEs based on sophisticated hardware protection mechanisms. Many TEEs have been made available to the public, such as enclave-based Intel SGX~\cite{sgx}, and VM-based Intel TDX~\cite{tdx}, AMD SEV SNP~\cite{amd-snp}. There also have been many works that leverage TEEs to build secure data sharing frameworks (e.g., ~\cite{squirrel}). Industries are actively exploring utilizing TEE-based solutions for secure data computation. For example, Google Research has proposed the \textit{Confidential Federated Computations} using TEE-based ledger systems~\cite{google-confidential}.

\section{Preliminary}
\label{sec:preliminary}

\vspace{0.25em}\noindent\textbf{Threat Model.}
\picachv's primary goal is to offer \textit{verifiable proofs} to data owners whose data is analyzed by any individual or group on the cloud inside TEEs that their data can be used only in accordance with the data use policies they specified. We thus assume that the TEE environment is completely safe. Attacks such as hardware side-channels, software vulnerabilities, or other exploits targeting TEEs are outside the scope of this paper and assumed to be mitigated by existing approaches (e.g., ~\cite{sectee}). The primary concerns are unintentional or intentional policy violations during data analytics. These may include improper data aggregation, inadequate anonymization, etc.

\vspace{0.25em}\noindent\textbf{Relational Algebra.} Relational algebra, a cornerstone of database theory, provides a set of operators for manipulating relational data.
Relational data is organized into tables, called relations, where each row represents a tuple (a single data entry) and each column represents an attribute (a property of the data). Relations are defined with a schema that specifies the attributes and their domains, ensuring consistency and structure in the data.
It forms the theoretical basis for relational database management systems and SQL. The fundamental operators include selection ($\sigma$) for filtering rows based on a condition, projection ($\pi$) for selecting specific columns, union ($\cup$) for combining tuples from two relations, set difference ($-$) for returning tuples in one relation but not another, Cartesian product ($\times$) for creating all possible tuple combinations from two relations, and the join ($\bowtie$) for concatenating tuples. By combining these operators, complex queries can be constructed to extract and transform data efficiently. Relational algebra's mathematical foundation ensures query precision and optimizability, making it crucial for database design and management, and beyond.
\section{Formalizing Data Use Policies}
\label{sec:formal}
In this section, we present the formalization of privacy policies so that they can accommodate policies written in human natural languages.
We first define a security lattice that provides a formal framework for expressing these policies.
Then we introduce the integration of data use policies with declassification requirements.


\subsection{Sensitivity Labels}
In reality, protecting data privacy often involves hierarchical sensitivity levels. For example, a user's full address is more sensitive than their zip code, and their zip code is more sensitive than their county. Similarly, individual purchase records are more sensitive than aggregated sales data. As data undergoes transformations like aggregation or anonymization, its sensitivity typically decreases. We first define sensitivity levels as labels in \autoref{def:lattice}.

\begin{definition}[Security lattice]\label{def:lattice}
    The security labels for \textsc{Picachv} are defined as $\mathcal{L} = \{\mathbf{L}, \mathbf{N}, \mathbf{A}, \mathbf{T}, \mathbf{H} \}$ such that $ \mathbf{L} \sqsubseteq \mathbf{N} \sqsubseteq \mathbf{A} \sqsubseteq \mathbf{T} \sqsubseteq  \mathbf{H}$. The security lattice is defined as $\tuple{\mathcal{L}, \sqsubseteq, \sqcup, \sqcap}$, satisfying:
    \begin{itemize}\setlength\itemsep{0em}
        \item $\mathbf{L} \sqsubseteq \ell, \forall \ell \in \mathcal{L}$.
        \item $\ell \sqsubseteq \mathbf{H}, \forall \ell \in \mathcal{L}$.
        \item $\ell_1 \sqsubseteq \ell_2 \Longleftrightarrow \ell_1 \sqcup \ell_2 = \ell_2$.
        \item $\ell_1 \sqsubseteq \ell_2 \Longleftrightarrow \ell_1 \sqcap \ell_2 = \ell_1$.
    \end{itemize}
    Note that $\sqsubseteq$ is antisymmetric, meaning that $\ell_1 = \ell_2$ if and only if $\ell_1 \sqsubseteq \ell_2 \wedge \ell_2 \sqsubseteq \ell_1$.
\end{definition}


The lattice in \autoref{def:lattice} includes \textbf{T} for data transformation requirements, \textbf{A} for data aggregation requirements, and \textbf{N} for noise addition requirements. This security lattice is motivated by the varying sensitivity levels in data governed by privacy policies. Different data elements and their derivatives possess distinct levels of sensitivity. Unlike other works with only two sensitivity levels (high $\bf H$ and low $\bf L$)~\cite{ifc-nonmalleable, declassification, beyond-policy-enforce}, our lattice captures nuanced requirements for data use policies. The real-world examples will be detailed later to strengthen our design. We first present the security lattice as follows. These labels themselves solely, however, cannot represent the policy requirements by data owners. We introduce \textit{policies} in the following section.

\subsection{Policies}
In our framework, each data element (specifically, individual cells within each relation) is associated with a distinct \textit{declassification} policy. This policy specifies sanitization procedures that ensure compliance with the required privacy policies. Upon satisfying all stipulated conditions, the data becomes eligible for disclosure. These declassification policies are fundamentally built upon the previously established base security labels, which collectively form a lattice $\cal L$.

We have adapted and refined the definition presented in \cite{downgrading} to formalize this concept to align with our specific scenario. The formal definition is as follows:

\begin{equation}
    p ::= \mathbf{L} \mid \ell^{\cal O} \rightsquigarrow p,
\end{equation}
where the lowest label $\mathbf{L}$ in the lattice indicates that all policies are enforced, allowing the data to be freely disclosed. The expression $\ell^{\cal O} \rightsquigarrow p$ signifies that for a given security label $\ell \in \cal L$, the data must be used following the security label $\ell$, and only operations defined in $\cal O$ can downgrade the data to the next policy $p$. A subtle consequence here is that operations will now have their own sensitivity levels. For example, a summation operation should have $\bf A$ because it is doing an aggregation.

We assume that we can always define a subset relation $\subseteq$ over $\cal O$, which can be either a vanilla set, or some complex structures like aggregation with group size requirements. For example, for aggregates, we can define $\mathcal{O} \mid n$ where $n$ denotes the minimum required group size, and $\cal O$ denotes the set of allowed aggregate operations. Additionally, given that the security lattice is now refined with operations, we present the refined lattice ``flows to'' relation in \autoref{fig:refined-lattice}. This relation is useful for composing policies, which will be detailed later in this section.

\begin{figure}[htbp]
    \centering
    \openup 2ex
    \begin{mdframed}
    \begin{gather*}
       \inference[\sc Lt]{\ell_1 \sqsubseteq \ell_2 \quad \ell_2 \not\sqsubseteq \ell_1}{\flowsto{\ell_1^{\mathcal{O}_1}} {\ell_2^{\mathcal{O}_2}}} \quad \inference[\sc Lt2]{\ell_1 = \ell_2 \quad \mathcal{O}_1 \subseteq \mathcal{O}_2}{\flowsto{\ell_1^{\mathcal{O}_1}} {\ell_2^{\mathcal{O}_2}}}
    \end{gather*}
    \end{mdframed}
    \caption{The refined lattice flows-to rules \fbox{$\flowsto{\ell_1^{\mathcal{O}_1}} {\ell_2^{\mathcal{O}_2}}$}.}
    \label{fig:refined-lattice}
\end{figure}
We present the declassification rules in \autoref{fig:policy-declassify}, modeled as a relation \(p \step{op, \ell} p^{\prime}\), where $op$ represents the operation applied to the policy, and $\ell$ denotes its sensitivity level. When an operation is applied to a policy, two cases arise: a) Operation sensitivity level is \textit{lower} than the policy’s sensitivity level: The absence of a corresponding reduction rule in \autoref{fig:policy-declassify} indicates an ill-formed case, resulting in a runtime error. b) Operation sensitivity level is \textit{greater} than or \textit{equal} to the policy’s sensitivity level: If the operation can successfully downgrade the policy, the policy is downgraded, as defined by the \textsc{Dlower} rule. Otherwise, the policy remains unchanged, as the required operation has not yet been applied, following the \textsc{DPreserve} rule.

\begin{figure}[htbp]
    \centering
    \openup 1ex
    \begin{mdframed}
    \begin{gather*}
        \inference[\sc DEmpty]{}{\mathbf{L} \step{op, \ell} \mathbf{L}} \quad
        \inference[\sc DLower]{p = \ell^{\mathcal{O}} \rightsquigarrow p' \quad op \in \mathcal{O}}{p \step{op, \ell} p'} \\
        \inference[\sc DPreserve]{p = \ell'^{\mathcal{O}} \rightsquigarrow p' \quad \ell' \sqsubseteq \ell \quad
        op \notin \mathcal{O}}{p \step{op, \ell} p}
    \end{gather*}
    \end{mdframed}
    \caption{The declassification rules \fbox{$p \step{op, \ell} p'$} for policy $p$.}
    \label{fig:policy-declassify}
\end{figure}

\noindentsec{Well-formed policy.} Specifically, because our policy focuses on declassification, policies must be ordered (in the sense of $\rightsquigarrow$) in the descending way. This reflects a practical basis: policies often specify the operations required to make data disclosable rather than to restrict it further. As shown in \autoref{fig:policy-wf}, a policy $p$ is well-formed if it consists of a chain where labels flow from higher to lower sensitivity levels, which can be verified through pattern matching on the refined ``flows to'' relation (\autoref{fig:refined-lattice}).

\begin{figure}[htbp]
    \centering
    \openup 2ex
    \begin{mdframed}
    \begin{gather*}
        \inference{}{\mathsf{wf}(\mathbf{L})} \quad \inference{}{\mathsf{wf}(\ell^{\cal O} \rightsquigarrow \mathbf{L})} \quad \inference{\mathsf{wf}(\ell_2^{O_2} \rightsquigarrow p) \\ \flowsto{\ell_2^{\mathcal{O}_2}}{\ell_1^{\mathcal{O}_1}}}{\mathsf{wf}(\ell_1^{\mathcal{O}_1} \rightsquigarrow \ell_2^{\mathcal{O}_2} \rightsquigarrow p)}
    \end{gather*}
    \end{mdframed}
    \caption{The well-formedness property of policy $p$.}
    \label{fig:policy-wf}
\end{figure}

\vspace{0.25em}\noindent\textbf{Policy Composition.} Policy composition is essential for scenarios where data from multiple sources converge. For instance, consider a medical dataset combining patient data from hospitals located in different states, such as California and Texas, each governed by its distinct data use policies. In cases where researchers aim to analyze the socioeconomic factors affecting patient health across states, they must adhere to the policies of \textit{both} datasets. This requires a mechanism for policy composition. \autoref{fig:policy-join} illustrates rules for a commutative composition process, ensuring policies are consistently merged. These rules ensure that joint analysis adheres to all policy requirements from both parties. Informally, policy join involves constructing a chain by inserting the appropriate policy label into the correct position based on the refined ``flows to'' relation. For policies sharing the same sensitivity levels in the \textsc{Merge} case, we apply the intersection operation $\cap$ to obtain the joined label $\ell^{\mathcal{O}_3}$.

\begin{figure}[htbp]
    \centering
    \openup 2ex
    \begin{mdframed}
    \begin{gather*}
        \inference[\sc JBotL]{}{\mathbf{L} \uplus p = p} \qquad \inference[\sc JBotR]{}{p \uplus \mathbf{L} = p} \\
        \inference[\sc Left]{\flowsto{\ell_2^{\mathcal{O}_2}}{\ell_1^{\mathcal{O}_1}} \quad p_1 \uplus (\ell_2^{\mathcal{O}_2} \rightsquigarrow p_2) = p_3}{\ell_1^{\mathcal{O}_1} \rightsquigarrow p_1 \uplus \ell_2^{\mathcal{O}_2} \rightsquigarrow p_2 = \ell_1^{\mathcal{O}_1} \rightsquigarrow  p_3} \\
        \inference[\sc Right]{\flowsto{\ell_1^{\mathcal{O}_1}}{\ell_2^{\mathcal{O}_2}} \quad (\ell_1^{\mathcal{O}_1} \rightsquigarrow p_1) \uplus  p_2 = p_3}{\ell_1^{\mathcal{O}_1} \rightsquigarrow p_1 \uplus \ell_2 \rightsquigarrow p_2 = \ell_2^{\mathcal{O}_2} \rightsquigarrow  p_3} \\
        \inference[\sc Merge]{\mathcal{O}_1 \cap \mathcal{O}_2 = \mathcal{O}_3 \quad p_1 \uplus p_2 = p_3}{\ell^{\mathcal{O}_1} \rightsquigarrow p_1 \uplus \ell^{\mathcal{O}_2} \rightsquigarrow p_2 = \ell^{\mathcal{O}_3} \rightsquigarrow p_3}
    \end{gather*}
    \end{mdframed}
    \caption{The composition rules for policy $p$.}
    \label{fig:policy-join}
\end{figure}



\subsection{Policy Expressiveness}
We argue that the declassification policies presented in this paper provide a practical abstraction, grounded in real-world scenarios. We provide three different privacy policy use cases to strengthen our argument.

\noindentsec{Scenario \#1: Medical Data Analysis.}
Medical data analysis serves as a concrete application of our policy design. Data collected by hospitals and medical research institutions is inherently sensitive, necessitating strict privacy regulations. To illustrate the applicability of our policies, we examined two examples: \textit{HIPAA Safe Harbor}\cite{hipaa}, a foundational standard for medical privacy, and the \textit{All-of-Us}\cite{allofus} project initiated by the U.S. National Institutes of Health (NIH) that extends the HIPPA boundary. For instance, under the HIPAA Safe Harbor framework, researchers must redact the last three digits of a zip code. This policy can be expressed as $\mathbf{T}^{\{ \rm redact(3) \}} \rightsquigarrow \mathbf{L}$. The \textit{All-of-Us} project mandates that aggregation must be performed on groups of at least 20 individuals, represented as  $\mathbf{A}^{\{ \rm \rm sum, avg, max, min ... \} \mid 20} \rightsquigarrow \mathbf{L}$.

\noindentsec{Scenario \#2: The U.S. Census Bureau.}
The U.S. Census Bureau’s 2020 disclosure avoidance guidance~\cite{us-census-bureau} employs differential privacy~\cite{dp} to add statistical noise to census data, preventing the identification of individuals in small demographic groups. The Census Bureau assigns varying privacy budgets $\varepsilon$ to different individuals and census blocks. \picachv's policy framework captures these requirements by associating $\mathbf{N}^{\Delta} \rightsquigarrow \mathbf{L}$ with the corresponding data objects. In this case, please note that \picachv leaves the task of specifying the noise to policymakers and does not dictate \textit{what} noise should be added to the corresponding data. Instead, it only ensures that a specific noise $\Delta$ is applied to the data.

\noindentsec{Scenario \#3: Policy for businesses.} An additional example illustrates how fine-grained policies can be applied in business contexts. 	Google provides users with extensive privacy rights, allowing them to control what information is collected and for what purposes\footnote{\url{https://safety.google/intl/en\_us/privacy/ads-and-data/\#controls}}. For example, consider a table where Google stores user demographic information. If users opt out of ad recommendations based on their age, a cell-level policy $\bf H^{\emptyset} \rightsquigarrow \bf L$ can be specific on the corresponding individuals of the attribute \texttt{age}, indicating that their ages can never be used.
Users can also specify other requirements like redaction, generalization, etc. that can be further applied on the base policy. If Google’s basic policy requires age data to be aggregated for groups of at least 100 people, policies can be “overlaid” to construct $\mathbf{T}^{\{redact\}} \rightsquigarrow \mathbf{A}^{\{ \rm sum, avg, \cdots \} \mid 100} \rightsquigarrow \mathbf{L}$.

\begin{figure*}[tb]
    \centering
    \begin{mdframed}
        \begin{subfigure}{0.45\textwidth}
        \fbox{Data model}
            \begin{align*}
            &\text{Type:} &\tau &\quad::= \mathtt{int \mid str \mid bool \mid} \\
            &\text{Schema:} &s &\quad::= \overline{\tau} \\
            &\text{Tuple:} &t &\quad::= \bot \mid \tuple{c, t} &\quad\\
            &\text{Cell:} &c &\quad::= \tuple{pv, id} \\
            &\text{Cell Identifier:} &id &\quad\in \mathbb{N} \\
            &\text{Tagged Value:} &v &\quad::= c \mid \overline{c}
            \end{align*}
        \fbox{Relational algebra}
            \begin{align*}
            &\text{Query:} &q &\quad::= R(id) \mid q_1 \cup q_2 \mid \pi_{\overline{e}}(q) \\ &\quad &\qquad &\quad \mid \quad q_1\bowtie_{\overline{n_1},\overline{n_2}} q_2 \mid \sigma_{\varphi}(q) \\
            &\quad &\qquad &\quad \mid \quad\gamma_{\overline{e} \mid \overline{g} \mid \varphi}(q) \\
            &\text{Expression:} &e, g, \varphi &\quad::= pv \mid \mathtt{col}(id) \mid \otimes e \\
            &\quad &\qquad &\quad \mid \quad e_1 \oplus e_2 \mid \mathcal{A}(e)
        \end{align*}
        \end{subfigure}%
        \begin{subfigure}{0.45\linewidth}
        \begin{align*}
            &\text{Primitive Value:} &pv &\quad\in \mathtt{str \mid bool \mid nums} \\
            &\text{Unary:} &\otimes &\quad::= \neg \mid - \mid + \mid ... \\
            &\text{Binary:} &\oplus &\quad::= \ge \mid \le \mid = \mid \neq ... \\
            &\text{Aggregate:} &\mathcal{A} &\quad::= \mathtt{sum \mid max \mid min \mid} ... \\
            &\text{Trace:} &tr &\quad::= \overline{\tuple{id, tt}} \\
            &\text{Trace type:} &tt &\quad::= \mathtt{TrNone}\ p \\
            &\quad &\qquad &\quad \mid \quad\mathtt{TrSingle}\ tt, op, p \\
            &\quad &\qquad &\quad \mid \quad \mathtt{TrMulti}\ \overline{tt}, op, p \\ 
            &\text{Policy:} &p &\quad \\
            &\text{Operator:} &op &\quad::= \oplus \mid \otimes \mid \mathcal{A} \\
            &\text{Group:} &G &\quad::= \tuple{t, \overline{n}}, n \in \mathbb{N} \\
            &\text{Data store:} &\Sigma &\quad::= \overline{\tuple{id, \tuple{R, \Gamma}}} \\
            &\text{Policy store:} &\Gamma &\quad::= \overline{\tuple{id, p}}
        \end{align*}
        \end{subfigure}
    \end{mdframed}
    \caption{The syntax for $\mathsf{RA^P}$.}
    \label{fig:ra-syntax}
\end{figure*}

\section{Formalizing \picachv}
\label{sec:ra}
This section formalizes the operational semantics of relational algebra ($\mathsf{RA}^{P}$), which supports PICACHV’s function as a runtime security monitor. The full syntax of relational algebra is defined in \autoref{fig:ra-syntax}. We first define an extended relational data model to support policy-associated data. 

\noindentsec{Data Model.} We begin by formalizing the underlying data model. The schema, $s$ is a list of primitive types, $\overline \tau$. A tuple $t$ is a dependent type based on the schema, with each element $t$ referred to as a \textit{cell}~\cite{verified-db}. A relation $R$ is a collection of tuples. We denote by $R[i]$ the $i$-th tuple in $R$, and slicing $R$ along indices $\overline{n}$ as $R \upharpoonright_{\overline{n}} T$. Each cell within a tuple is assigned a unique identifier ($id$) to enable policy lookup in the given environment detailed later. We visualize this in \autoref{fig:data-model}.

\noindentsec{Expressions.} In our model, expressions take five forms. Two atomic expressions include the primitive value $pv$, used as constants, and the column identifier $\mathtt{col}(id)$, for selecting values from a given column. Complex expressions include unary expressions $\otimes e$, binary expressions $e_1 \oplus e_2$, and aggregates $\mathcal{A}(e)$. Expressions are used as: 1) predicates in \texttt{select} for tuple filtering ($\varphi$) and \texttt{aggregate} ($g$) for group filtering, or 2) actual values evaluated as results.

\noindentsec{Relational algebra.} As illustrated in \autoref{fig:ra-syntax}, a query $q$ includes a special operator for data retrieval and five fundamental relational operators. A relation $R$ indexed by its identifier $id \in \mathbb{N}$ is denoted as $R(id)$. The union operator $q_1 \cup q_2$ merges two subqueries $q_1$ and $q_2$ that share the same schema. The join operator $q_1 \bowtie_{\overline{n_1}, \overline{n_2}} q_2$ combines results from two subqueries $q_1$ and $q_2$ based on columns specified by two lists of indices $\overline{n_1}$ and $\overline{n_2}$. The projection operator $\pi_{\overline{e}}(q)$ evaluates expressions $e \in \overline{e}$ on every row in the result of the query $q$. The selection operator $\sigma_{\varphi}(q)$ filters rows of $q$ that satisfies the predicate $\varphi$. The aggregate operator $\gamma_{\overline{e} \mid \overline{g} \mid \varphi}(q)$ is slightly more complex. It computes aggregates $\overline{e}$, groups rows by $\overline{g}$, and filters groups satisfying $\varphi$ (i.e., the \texttt{having} clause). Additionally, it specifies lists of aggregate expressions $\overline{e}$. Here, the group information $G$ includes the active tuple $t$, which represents the grouped keys (a list of column indices used for grouping), and the group-by indices $\overline{n}$ corresponding to the tuples in $R$ within the current group. For example, consider again the relation shown in \autoref{fig:data-model}, where the grouping is performed on the $\tau_1$ column. Hence, the grouping key is $\tau_1$, and there will be two groups whose representative values are $v_1$ and $v_2$, respectively. In this case, the first group $G_1$ would be $\tuple{\tuple{v_1, \bot}, \{0, 1\} }$, and the second one $G_2$ would be $\tuple{\tuple{v_2, \bot}, \{2\} }$, assuming that the index of each row starts at 0.


\begin{figure}[htbp]
    \centering
\resizebox{0.95\linewidth}{!}{\begin{tikzpicture}[
    table cell/.style={draw, minimum width=2.25cm, minimum height=0.75cm, align=center},
    header cell/.style={draw, fill=gray!30, text=black, minimum width=2.25cm, minimum height=0.75cm, align=center},
    label/.style={font=\small, inner sep=1pt},
    arrow/.style={-{Stealth}, thick},
    brace/.style={decorate, decoration={brace, amplitude=8pt}, thick},
    dashed box/.style={draw, thick, dashed, inner sep=4pt},
]

\node[header cell] (h1) at (0, 0) {$\bf \tau_1$};
\node[header cell] (h2) [right=0cm of h1] {$\bf \tau_2$};
\node[header cell] (h3) [right=0cm of h2] {$\bf \tau_3$};

\node[table cell] (r11) [below=0cm of h1] {$(v_1, id_1)$};
\node[table cell] (r12) [below=0cm of h2] {$(v_3, id_2)$};
\node[table cell] (r13) [below=0cm of h3] {$(v_3, id_3)$};

\node[table cell] (r21) [below=0cm of r11] {$(v_1, id_4)$};
\node[table cell] (r22) [below=0cm of r12] {$(v_5, id_5)$};
\node[table cell] (r23) [below=0cm of r13] {$(v_6, id_6)$};

\node[table cell] (r31) [below=0cm of r21] {$(v_2, id_7)$};
\node[table cell] (r32) [below=0cm of r22] {$(v_8, id_8)$};
\node[table cell] (r33) [below=0cm of r23] {$(v_9, id_9)$};

\node[label, right=.5cm of r13.east] (tuple) {tuple $t$};
\draw[brace] ([yshift=0.3cm]h1.west|-h1.north) -- ([yshift=0.3cm]h3.east|-h1.north) node[midway, above=0.3cm] {schema $s$};
\draw[brace] ([xshift=-0.3cm]r31.west|-r31.south) -- ([xshift=-0.3cm]r11.west|-r11.north) node[midway, left=0.4cm] {relation $R$};

\draw[arrow] (r13.east) -- (tuple.west);

\draw[dashed box] ([xshift=-0.2cm, yshift=0.2cm]r11.north west) ([xshift=0.2cm, yshift=-0.2cm]r33.south east);
\draw[dashed box] ([xshift=-0.2cm, yshift=0.2cm]h1.north west) ([xshift=0.2cm, yshift=-0.2cm]h3.south east);

\end{tikzpicture}}
    \caption{The relational model. 
    }
    \label{fig:data-model}
\end{figure}

\noindentsec{Program trace.} The program trace $tr$ serves two purposes when performing expression evaluations which be introduced next. First, it ensures that all data undergoes the appropriate operations before being released by recording the history of operations ($op$) performed on each cell. Second, it maintains the current policy $p$ associated with the cell being evaluated. Accordingly, $tr$ is represented as a list of tuples containing cell identifiers and their corresponding trace types $t$. The intuition behind the definition of $t$ is straightforward: within the five relational operators, a cell may undergo one of two types of transformations. These transformations involve either a single data source, as in the case of $\mathtt{TrSingle}$ caused by \texttt{project}s, or multiple data sources, as in $\mathtt{TrMulti}$, typically resulting from operations like \texttt{aggregate}s or \texttt{join}s. For cells that have not undergone any operations, their trace type is represented as $\mathtt{TrNone}\ p$, with $p$ as policyies associated with the cell.




\subsection{Formal Semantics}
This section introduces the core reduction rules for $\mathsf{RA}^{P}$, starting with the rules for evaluating expressions in relational algebra, followed by the rules for relational operators.

\subsubsection{Expressions}
\begin{figure*}[tb]
    \centering
    \begin{mdframed}
    \openup 1ex
    \begin{gather*}
        \inference[\sc Column]{T = t \quad n < |t|}{\tuple{tr, T}\ \textcolor{TealBlue}{\xrightarrow{\mathtt{col}(n)}} \ \tuple{tr, t[n]}} \qquad
        \inference[\sc ColumnAgg]{T = \overline{t} \quad \forall i \in |\overline{t}| \Longrightarrow
        n < |t_i|}{\tuple{tr, T}\ \textcolor{TealBlue}{\xrightarrow{\mathtt{col}(n)}}\ \tuple{tr, \bigcup_{i} t_i[n]}} \\
        \inference[\sc Aggregate]{\tuple{tr, T} \ \textcolor{TealBlue}{\xrightarrow{e}}\ \tuple{tr'', v} \quad v = \overline{\tuple{pv, id}} \\ \tuple{tr'', t}\ \textcolor{TealBlue}{\xrightarrow{\mathtt{agg}, v}}\ \tuple{tr', v'}}{\tuple{tr, T}\ \textcolor{TealBlue}{\xrightarrow{\mathtt{agg}(e)}}\ \tuple{tr', v'}} \\
        \inference[\sc Unary]{\tuple{tr, T}\ \textcolor{TealBlue}{\xrightarrow{e}} \tuple{tr'', v}\ \quad v = \tuple{pv, id} \\
        \tuple{tr'', T}\ \textcolor{TealBlue}{\xrightarrow{\otimes, v}}\ \tuple{tr', v'}
        }{\tuple{tr, T}\ \textcolor{TealBlue}{\xrightarrow{\otimes\ e}}\ \tuple{tr', v'}} \qquad
        \inference[\sc UnaryAgg]
        {\tuple{tr, T}\ \textcolor{TealBlue}{\xrightarrow{e}}\ \tuple{tr'', v} \quad v = \overline{\tuple{pv, id}} \quad \\
        \forall i \in |v| \Longrightarrow \left(\tuple{tr'', T}\ \textcolor{TealBlue}{\xrightarrow{\otimes, v}}\ \tuple{tr_i', v_i'}\right)}
        {\tuple{tr, T}\ \textcolor{TealBlue}{\xrightarrow{\otimes\ e}}\ \tuple{\bigcup_{i} tr_i',\ \bigcup_{i} v_i'}} \\
        \inference[\sc Binary]{\tuple{tr, T} \ \textcolor{TealBlue}{\xrightarrow{e_1}}\ \tuple{tr_1, v_1} \quad \tuple{tr, T}\ \textcolor{TealBlue}{\xrightarrow{e_2}}\ \tuple{tr_2, v_2} \\
        v_1 = \tuple{pv_1, id_1} \quad v_2 = \tuple{pv_2, id_2} \\
        \tuple{tr_1 \cup tr_2, T}\ \textcolor{TealBlue}{\xrightarrow{\oplus, v_1 :: v_2 :: \mathtt{nil}}}\  \tuple{tr', v'}}{\tuple{tr, T}\ \textcolor{TealBlue}{\xrightarrow{e_1 \oplus e_2}}\ \tuple{tr', v'}} \qquad
        \inference[\sc BinaryAgg]{\tuple{tr, T} \ \textcolor{TealBlue}{\xrightarrow{e_1}} \tuple{tr_1, v_1}\ \quad \tuple{tr, T}\ \textcolor{TealBlue}{\xrightarrow{e_2}}\ \tuple{tr_2, v_2} \\
         v_1 = \overline{\tuple{pv_1, id_1}} \quad v_2 = \overline{\tuple{pv_2, id_2}} \quad |v_1| = |v_2| \\
        \forall i \in |v_1| \Longrightarrow \left(\tuple{tr_1 \cup tr_2, T}\ \textcolor{TealBlue}{\xrightarrow{e_1 \oplus e_2}}\ \tuple{tr_i, v_i'} \right)}{\tuple{tr, T}\ \textcolor{TealBlue}{\xrightarrow{e_1 \oplus e_2}}\ \tuple{\bigcup_{i} tr_i', \bigcup_{i} v_i'}}
    \end{gather*}
    \end{mdframed}
    \caption{Main expression evaluation rules for \fbox{$\tuple{tr, T} \ \textcolor{TealBlue}{\overset{e}{\longrightarrow}}\ \tuple{tr', v}$}.}
    \label{fig:exp-semantics}
\end{figure*}
Expressions form the foundation of relational algebra. Expression evaluation is defined as the relation\footnote{In Coq, an additional argument for the maximum allowed steps must be introduced to satisfy the termination checker.}:
\[
    \tuple{tr, T}\ \textcolor{TealBlue}{\overset{e}{\longrightarrow}}\ \tuple{tr', v},
\]
where a trace $tr$ and a tuple context $T ::= t \mid \overline{t}$ (represents either a single tuple $t$ or a list of tuples in an aggregate context) are inputs, and the evaluation of the expression $e$ produces a value $v$ along with an updated trace $tr^{\prime}$. To highlight expression evaluation that involves policy transformation, we define an auxiliary expression evaluation relation as follows.

\[
    \tuple{tr, T}\ \textcolor{TealBlue}{\overset{f, v}{\longrightarrow}}\ \tuple{tr', v'},
\]
where $f$ is the function being applied to tagged value $v$. We present the main expression evaluation rules in \autoref{fig:exp-semantics}, where we categorize expression evaluation based on the tuple context $T$. The \textsc{Column} rule is straightforward, as it simply retrieves the value at the specified index $n$ from the tuple $t$. For \textsc{Unary} and \textsc{Binary} expressions, the rules invoke the corresponding auxiliary evaluation functions. In the aggregate context, the evaluation works similarly but applies the rules iteratively to each element in $v$.

\begin{figure}[h]
    \centering
    \small
    \begin{mdframed}
    \openup 1ex
    \begin{gather*}
        \inference{v = \tuple{pv, id} \quad \tuple{id, p} \in tr \\  \text{\colorbox{gray!30}{$p \step{f, \mathbf{T}} p'$}} \quad tr' = tr[\tuple{id, p} \mapsto \tuple
        {id, p'}]
        }{\tuple{tr, T} \ \textcolor{TealBlue}{\xrightarrow{f, v}}\ \tuple{tr', \tuple{\denote{f}(pv), id}}} \tag{\sc FUnary}\\
        \inference{ \tuple{id_1, p} \in tr \quad \tuple{id_2, \mathbf{L}} \in tr \quad id = \mathtt{new\_id}(tr) \\
        v = \tuple{pv_1, id_1} :: \tuple{pv_2, id_2} :: \mathtt{nil} \\ \text{\colorbox{gray!30}{$p \step{f(\cdot, pv_2), \mathbf{T}} p'$}} \quad tr' = tr[\tuple{id, p} \mapsto \tuple
        {id, p'}]}{\tuple{tr, T}\ \textcolor{TealBlue}{\xrightarrow{f, v}}\ \tuple{tr', \tuple{\denote{f}(pv_1, pv_2), id}}} \tag{\sc FBinary} \\
        \inference{v = \overline{\tuple{pv, id}} \quad
        pvs = \mathtt{map}(\mathtt{fst}, v) \quad ids = \mathtt{map}(\mathtt{snd}, v)  \\
        \text{\colorbox{gray!30}{$\forall i \in |P| \Longrightarrow \left(\tuple{ids_i, p_i} \in tr\right)\wedge \left(p_i \step{\mathtt{agg}, \mathbf{A}} p_i'\right)$}} \\
        tr' = \tuple{id', \biguplus_{i} p_i'} :: tr \quad id' = \mathtt{new\_id}(tr) \quad P = \{ p_1, \cdots \} 
        }{\tuple{tr, T}\ \textcolor{TealBlue}{\xrightarrow{\mathtt{agg}, v}}\ \tuple{tr', \tuple{\mathtt{fold}(\denote{\mathtt{agg}}, pvs), id'}}} \tag{\sc FAgg}
    \end{gather*}
    \end{mdframed}
    \caption{Auxiliary rules for \fbox{$\tuple{tr, T}\ \textcolor{TealBlue}{\overset{f, v}{\longrightarrow}}\ \tuple{tr', v'}$} that directly manipulates policy checks and transitions that are \colorbox{gray!30}{highlighted}.}
    \label{fig:exp-semantics-downdown}
\end{figure}

The auxiliary rules in \autoref{fig:exp-semantics-downdown} are more interesting. The \textsc{FUnary} rule governs unary function application, where the program trace $tr$ must contain the policy $p$ associated with the tagged value $v$ (which includes a unique identifier $id$). The intended operation $f$ is applied to $p$, transitioning the label (\colorbox{gray!30}{highlighted}) and producing an updated policy $p’$ (if possible as defined in the \autoref{fig:policy-declassify}). This change is reflected in the updated program trace as $tr[\tuple{id, p} \mapsto \tuple{id, p’}]$. The function $f$ is then interpreted as $\denote{f}$ and applied to the primitive value $pv$ carried by $v$, yielding $\denote{f}(pv)$. The \textsc{FBinary} rule follows a similar process but includes an additional constraint: the second argument $v_2$ must be clean, meaning its policy must be $\mathbf{L}$. The \textsc{Aggregate} rule is slightly more complex. Before applying the operation, it ensures that all policies in the list of primitive values are compatible with the aggregate function. It then transitions the labels to obtain each updated policy $p_i’$. The final result is computed by folding over the value list $pvs$, with the resulting policy label being the composition of all $p_i’$. We use $\mathtt{map}$ to split $v$ into two lists using the pair projection functions $\mathtt{fst}$ and $\mathtt{snd}$: one list for values, $pvs$, and another for identifiers, $ids$. It has type:
\begin{align*}
    \mathtt{map}: (A \to B) \to \overline{A} \to \overline{B},
\end{align*}
which applies a function $f: A \to B$ on each element of the list $\overline{A}$, producing a list of results $\overline{B}$. Next, the list of primitive values and their identifiers is processed to compute a result using the \texttt{fold} operation, which has the type:
\begin{align*}
    \mathtt{fold}: (B \to A \to B) \to B \to \overline{A} \to B,
\end{align*}
This operation takes a higher-order function $f$ (which accumulates elements of the list), an identity element $B$, a list of elements $\overline{A}$, and produces a final result $B$. Afterward, a new identifier for the result is generated by the \texttt{new\_id} function\footnote{We assume that this function is like a UUID generator so we do not need to worry about id conflicts.}.

\begin{figure*}[t]
    \centering
    \begin{mdframed}
    \openup 1ex
    \small
    \fbox{$\tuple{t, R, \overline{n_1}, \overline{n_2}} \downarrow^{tr}_{tr'} R'$}
    \begin{equation*}
        \inference[\sc JoinT]{t \upharpoonright_{\overline{n_1}} \tuple{t_1, t_1^{\star}} \quad P_1 = \{ p_{1_1}, \cdots \} \quad \forall id \in \mathtt{ids}(t_1) \Longrightarrow \tuple{id_{1_i}, p_{1_i}} \in tr \quad tr = \bigcup tr_i \\
        \forall i \in |R| \Longrightarrow 
        \left(\text{$\begin{aligned}
            &R[i] \upharpoonright_{\overline{n_2}} \tuple{t_2, t_2^{\star}} \quad P_2 = \{ p_{2_2}, \cdots \} \quad \forall id \in \mathtt{ids}(t_2) \Longrightarrow \tuple{id_{2_i}, p_{2_i}} \in tr \\ 
            &\tuple{t_i, tr_i} =
            \begin{cases}
            \tuple{t_1^{\star} \parallel t_1' \parallel t_2^{\star}, tr'}, \quad &\text{\bf if $t_1 \doteq t_2$}\\
            \tuple{\bot, tr}, \quad &\text{\bf otherwise}
        \end{cases} \quad t_1' = \tuple{id_1, pv, \tuple{\cdots}} \quad \forall j \in |t_1'| \Longrightarrow \tuple{id_1, \text{\colorbox{gray!30}{$p_1 \uplus p_2$}}} \in tr'
        \end{aligned}$}\right)
        }{\tuple{t, R, \overline{n_1}, \overline{n_2}} \downarrow^{tr}_{tr'} R'}
    \end{equation*}
    \fbox{$\Sigma \vdash q \ \textcolor{blue}{\Downarrow}\ \tuple{R, tr}$}
    \begin{gather*}
        \inference[\sc Join]{
        \Sigma \vdash q_1  \ \textcolor{blue}{\Downarrow}\  \tuple{R_1, tr_1} \quad \Sigma \vdash q_2 \ \textcolor{blue}{\Downarrow}\ \tuple{R_2, tr_2} \quad tr' = tr_1 \cup tr_2 \\
        \forall i \in |R_1| \Longrightarrow \left(\tuple{R[i], R_2, \overline{n_1}, \overline{n_2}} \downarrow^{tr'}_{tr_i} R_i \right)
        }{\Sigma \vdash \left(q_1 \bowtie_{\overline{n_1}, \overline{n_2}} q_2\right) \ \textcolor{blue}{\Downarrow}\ \tuple{\bigcup_{i} R_i, \bigcup tr_i}}  \qquad
        \inference[\sc Union]{\Sigma \vdash q_1 \ \textcolor{blue}{\Downarrow}\ \tuple{R_1, tr_1} \quad \Sigma \vdash q_2 \ \textcolor{blue}{\Downarrow}\ \tuple{R_2, tr_2}}{\Sigma \vdash \left(q_1 \cup q_2 \right) \ \textcolor{blue}{\Downarrow}\ \tuple{R_1 \cup R_2, tr_1 \cup tr_2}} \\
        \inference[\sc Select]{\Sigma \vdash q \ \textcolor{blue}{\Downarrow}\ \tuple{R, tr} \quad \forall i \in |R| \Longrightarrow \tuple{\star, R[i]} \ \textcolor{TealBlue}{\xrightarrow{\varphi}}\  \tuple{\star, b_i})\\
        }{\Sigma \vdash \sigma_{\varphi}(q) \ \textcolor{blue}{\Downarrow}\ \tuple {R \times ( b_1, \cdots )^{T}, tr}} \qquad  \inference[\sc Relation]{\tuple{id, \tuple{R, \Gamma}} \in \Sigma \\ tr = \mathtt{map}(\Gamma, \lambda x. \tuple{\mathtt{fst}(x), \mathtt{TrNone}\ \mathtt{snd}(x)})}{\Sigma \vdash R(id) \ \textcolor{blue}{\Downarrow}\ \tuple{R, tr } } \\
        \inference[\sc Project]{\Sigma \vdash q \ \textcolor{blue}{\Downarrow}\ \tuple{R', tr'} \quad \forall i \in |\overline{e}|, \forall j \in |R'| \Longrightarrow \left(\tuple{tr', R'[j]}\ \textcolor{TealBlue}{\xrightarrow{e_i}}\ \tuple{tr_{ij}, v_{ij}} \quad R[j] = ||_{i} \ v_{ij} \right)
        }{\Sigma \vdash \pi_{\overline{e}}(q) \ \textcolor{blue}{\Downarrow}\ \tuple{\bigcup_{j} R[j], \bigcup_{i, j} tr_{ij}}} \\
        \inference[\sc Aggregate]{\Gamma \vdash q \ \textcolor{blue}{\Downarrow}\ \tuple{R', tr'} \quad R' \searrow_{\overline{g}} \overline
        G \\ 
        \forall i \in |\overline{e}|, \forall j \in |\overline{G}| \Longrightarrow \left( G_j = \tuple{t_j, \overline{n}_j} \quad R' \upharpoonright_{\overline{n}_i} T \quad \tuple{tr', T}\ \textcolor{TealBlue}{\xrightarrow{e_i}}\ \tuple{tr_{ij}, v_{ij}} \quad R[j] = ||_{i}\ (t_i || v_{ij}) \quad \tuple{\star, t_j}\ \textcolor{TealBlue}{\xrightarrow{\varphi}}\ \tuple{\star, b_j} \right)}
        {\Sigma \vdash \gamma_{\overline{e} \mid \overline{g} \mid \varphi}(q) \ \textcolor{blue}{\Downarrow}\ \tuple{\left(\bigcup_{j} R[j] \right) \times \{b_j, \cdots \}^{T}, \bigcup_{i, j} tr_{ij}}}
    \end{gather*}
    \end{mdframed}
    \caption{Selected reduction rules for $\mathsf{RA^P}$.}
    \label{fig:rap-semantics}
\end{figure*}

\noindentsec{Trusted Blackbox Functions.} In the expression evaluation rules defined in \autoref{fig:rap-semantics}, we treat functions as \textit{blackboxes}~\cite{privguard} for two primary reasons. First, no limitations are imposed on function implementations, allowing library developers to create functions or user-defined functions (UDFs) in any programming language and style. This flexibility, however, complicates policy enforcement, as it requires not only understanding the semantics of these functions but also supporting a wide array of programming languages. Second, it is more practical and secure to establish a barrier between third-party code and \textsc{Picachv}. This separation allows for better control and monitoring of interactions between external functions and the core system. We also believe that programmers can provide vetted code for these functions.







\subsubsection{Relational Operators}
We now turn our attention to the reduction rules for the operators of our core calculus $\mathsf{RA^P}$, as illustrated in \autoref{fig:rap-semantics}. This behavior is formalized using big-step operational semantics, represented by the following judgment form:

\[
\Sigma \vdash q \ \textcolor{blue}{\Downarrow}\ \tuple{R, tr}
\]

This establishes a relation between the initial evaluation context $\Sigma, q$, which consists of the data store and the query $q$, and the resulting relation $R$ along with the updated trace $tr$.

We present reduction rules for relational operators in \autoref{fig:exp-semantics}. The \textsc{Join} rule first evaluates its sub-queries, yielding their results $R_1$ and $R_2$. It then iterates over the tuples in the left relation, such that for each tuple $t_1 \in R_1$, it attempts to concatenate $t_1$ with all tuples $t_2 \in R_2$ from the right relation. The \textsc{JoinT} rule, which is used by the \textsc{Join} rule, is particularly noteworthy. This rule attempts to join a tuple $t$ with a relation $R$. Since a join operation requires specifying which columns are used as keys, we use $t \upharpoonright_{\overline{n}} \tuple{t_1, t_1^{\star}}$ to denote splitting the tuple $t$ into two parts: $t_1$, which contains the selected columns (keys), and $t_1^{\star}$, which contains the remaining columns. The \textsc{JoinT} rule then iterates over the tuples in the right relation $R$. It ensures that all identifiers and policies of the tuple being joined are present in the initial trace $tr$. During each iteration, if the tuples $t_1$ and $t_2$ agree on the joined part (denoted as $t_1 \doteq t_2$), their policies $p_1$ and $p_2$ are composed as $p_1 \uplus p_2$, and the new policy is inserted into the updated trace $tr’$. The \textsc{Union} rule simply unions the result of the two sub-queries.

The base rule is presented in \textsc{Relation} where we fetch the policy from the policy store and transform it into a trace $tr$ consisting of \texttt{TrNone}. For \textsc{Select}, we iterate over each tuple $R[i] \in R$ obtained from the evaluated result of the subquery, and we evaluate the predicate $\varphi$ thereon. Since evaluating the predicate does not involve any policy-related operations, we disregard the program trace, using $\star$ as a special placeholder to indicate that it is irrelevant in this context and can be any well-typed traces. The evaluation of $\varphi$ on each tuple produces a boolean value, resulting in a vector $\{ b_i, \ldots \}$. This vector is then used to filter the relation $R$ through the operation $\times$, retaining only the tuples that satisfy the predicate. \textsc{Project} evaluates each expression $e_i \in \overline{e}$ on every tuple $R’[j] \in R’$. The results are concatenated into a single tuple $R[j]$, and all such tuples are combined through a union operation to produce the final result. The \textsc{Aggregate} rule begins by deriving the grouping information $R’ \searrow_{\overline{g}} \overline{G}$ from $R’$, which is obtained from the evaluation result of the sub-query. It then iterates over the list of aggregate expressions, evaluating each expression $e_i$ using the tuple context $T$, which is created by slicing $R’$ based on $\overline{n}_j$. Eventually, we obtain the result by applying the grouping predicate on the intermediate result.

\subsection{Security Conditions}

\begin{figure}[h]
    \centering
    \small
    \begin{mdframed}
    \begin{gather*}
        \inference{}{\mathcal{E}(\mathtt{TrNone}\ \ell) = \ell}
        \qquad 
        \inference{T = \left(\biguplus_{t_i \in \mathcal{E}(t)} (t_i \step
        {op} \ell)\right)}{\mathcal{E}(\mathtt{TrSingle} \ t, op, \ell) = T} \\
        \inference{T = \left(\biguplus_{t_i \in \overline{t}}\mathcal{E} (t_i)\right)}{\mathcal{E}(\mathtt{TrMulti}\ \overline{t}, op, \ell) = \left(\biguplus_{t \in T} (t \step{op} \ell) \right)}
    \end{gather*}
    \end{mdframed}
    \caption{Trace extraction rules.}
    \label{fig:trace-extraction}
\end{figure}

The key security property of \textsc{Picachv}'s semantics is ensuring that the enforcement is \textit{sound}. Following this work~\cite{downgrade-non-inter}, the enforcement mechanism ensures that data tagged as low has been downgraded via specified functions in accordance with its policy.

\begin{definition}[Relaxed non-interference]
    For any data store $\Sigma$, query $q$, either the evaluation of $\Sigma, q$ results in an error, i.e., $\Sigma \vdash q \ \textcolor{blue}{\Downarrow}\ \text{\ding{55}}$, or the following holds:
    \begin{align*}
        &\Sigma \vdash q \ \textcolor{blue}{\Downarrow}\ \tuple{R, tr} \\
        &\qquad \Longrightarrow
        (\forall c \in \denote{R}, \tuple{c, \mathbf{L}} \in tr \Longrightarrow \mathcal{E}(c) \approx \Sigma(c)),
    \end{align*}
where $\denote{R}$ means to extract identifiers of all data in the relation, $\cal E$ is a trace extraction function defined in \autoref{fig:trace-extraction}, $\approx$ is a compatible relation between policies, and we use $\Sigma(c)$ to denote the initial policy for $c$. Informally, $\cal E$ identifies every transformation path of the data involved in computing $c \in R$. We define the compatible relation in \autoref{fig:comp}.
\end{definition}

\begin{figure}[h]
    \centering
    \begin{mdframed}
    \begin{equation*}
        \inference{}{p \approx p} \qquad \inference {op \in \mathcal{O} \quad \mathsf{wf}(p) \quad p_1 \approx p_2 }{(p \step{op, \ell} p_1) \approx (\ell^{\cal O} \rightsquigarrow p_2)}
    \end{equation*}
    \end{mdframed}
    \caption{Rules for the compatible relation $\approx$}
    \label{fig:comp}
\end{figure}

This security definition encapsulates two key scenarios: (1) evaluation results in an error due to a policy breach, returning no output. (2) Relaxed non-interference ensures that all data tagged as low $\mathbf{L}$ has undergone proper declassification procedures as specified by the data owner. Notably, when no declassification is allowed (i.e., all data is set to $\bf H$), this definition reduces to standard non-interference. For instance, consider a simple program that performs an aggregation over the \texttt{age} attribute with a policy $\mathbf{A}^{\bf avg} \rightsquigarrow \mathbf{L}$, oututting a final relation $R$ containing a single value, $\tuple{pv, id}$. Then $\mathcal{E}(id) \equiv (\mathcal{E}(c_0) \step{\bf avg, A} \mathbf{L}) :: (\mathcal{E}(c_1) \step{\bf avg, A} \mathbf{L}) :: \cdots$.

\begin{figure*}[t]
    \centering
    \includegraphics[width=\linewidth]{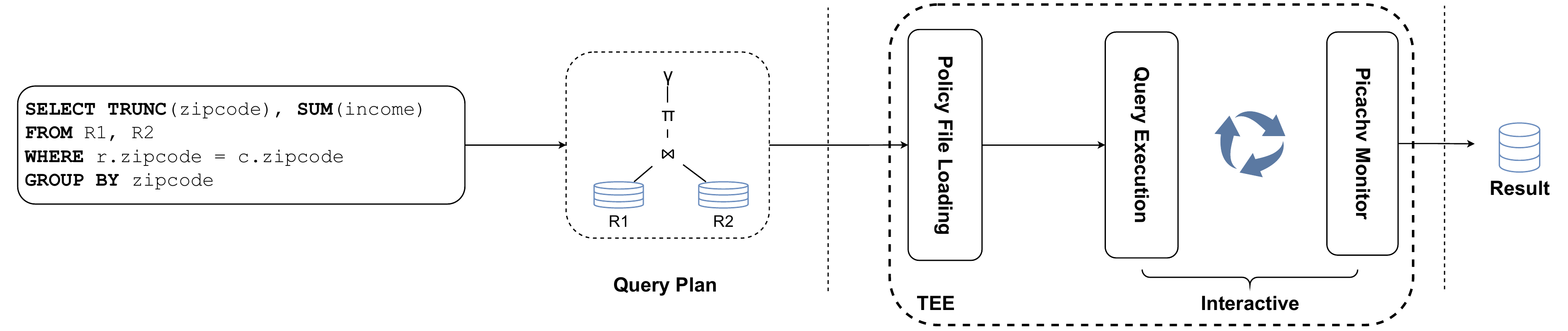}
    \caption{The high-level query execution workflow where we put the query execution engine and \picachv runtime monitor inside a TEE.}
    \label{fig:system-overview}
\end{figure*}

\begin{theorem}[Soundness]\label{thm:soundness}
    The semantics enforces relaxed non-interference.
\end{theorem}

\begin{theorem}[Strict non-interference]\label{thm:non-inter}
    If no declassification is permitted, then our semantics enforces strict non-interference.
\end{theorem}

%
\section{Implementation}
\label{sec:implementation}
In this section, we explain how the \picachv runtime monitor operates and how policies are encoded.
\subsection{Implementing Policies}
\noindentsec{Encoding policies using shadow tables.} As described in \autoref{sec:formal}, each relation is modeled as a list of tuples, with each cell assigned a unique identifier $id$ linking it to its corresponding policies. To address the inefficiency of frequent policy lookups during execution, particularly in parallel processing environments (that require locks), we introduce the concept of \textit{shadow tables}, which maintains only the policy tags for the given relation in a one-to-one correspondence. Shadow tables mirror the structure of the original tables but store policy tags instead of the actual data. This design separates data from its policies, enabling efficient runtime policy enforcement. This way, the policies for a given relation are stored separately with the data and will be loaded at runtime when query is being executed. In our implementation, shadow tables that store policies are encoded in Apache Parquet format where we serialize policies into byte arrays and compress them. 

\noindentsec{Support for flexible policies.} Shadow tables also support flexibility by allowing policymakers to load different policies for the same relation from separately stored policy files. For instance, researchers from different institutions working on the same dataset may require distinct data use policies tailored to their roles. The policymaker who controls the data can design different policies for the same patient data depending on the roles of the researcher. At the same time, this way allows policies to be arbitrarily composed when multiple relations with different policy requirements are being joined during analysis. Even better is that this design allows for policy overlay. Imagine if we have a ``base policy'' for a given table and there would be several users having different privacy preferences. As such, these requirements can be overlayed on the base policy by applying the policy composition rules shown in \autoref{fig:policy-join}.

\subsection{Implementing \picachv}
At a high level, the implementation of \picachv consists of two key components: a formal proof written in Coq and a runtime monitor developed in Rust, designed as a standalone dynamic library. This library can be integrated into existing data analytics frameworks through foreign function interfaces (FFIs), requiring minimal modifications to the codebase. Frameworks like \texttt{Pandas} and \texttt{SparkQL} can invoke \picachv’s monitoring APIs to enforce data use policies dynamically whenever an executor is executed.

The high-level workflow of \picachv is illustrated in \autoref{fig:system-overview}. The process begins with the parsing and transformation of a query into a query plan, which is subsequently forwarded to the execution engine. \picachv identifies and retrieves the relevant policy files from disk, based on the tables referenced in the query—such as R1 and R2 in this example. Query execution proceeds interactively, with \picachv actively monitoring each node’s execution in real time. If the query adheres to the specified policies, the computed results are allowed to exit \picachv. Conversely, if the query violates any policies, an error is raised to block non-compliant results from being returned. Note that we place \picachv and data inside TEEs for confidentiality, integrity, and verifiability.

\begin{figure}[h]
    \centering
    \includegraphics[width=\linewidth]{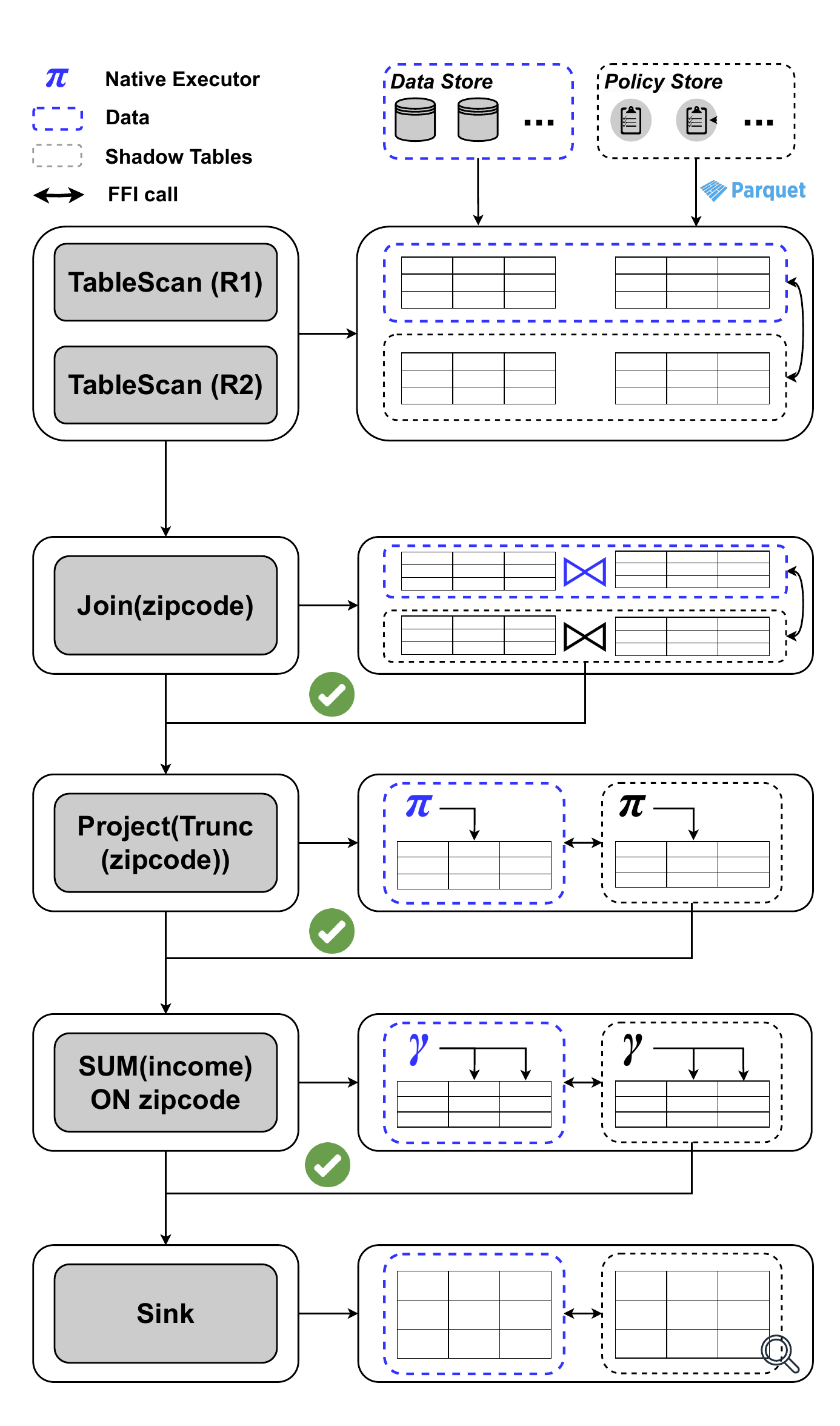}
    \caption{The query execution phase.}
    \label{fig:query-execution}
\end{figure}

\noindentsec{Query execution.} \autoref{fig:query-execution} details the query execution process in \picachv. Data and policy shadow tables are first fetched by \texttt{TableScan} operator, then processed by \texttt{Project} and \texttt{Aggregate}, and finally \texttt{sink} is applied. At a high level, since shadow tables are maintained for the original tables, the execution process is divided into two parallel phases. Recall that the way shadow tables are implemented can support flexible policies for the \textit{same table} under different circumstances. The effects of relational operators on the shadow tables will being actively captured by the semantics described in \autoref{fig:rap-semantics}, while the actual data is processed using the native query executor (in {\color{blue}{blue}}). In other words, the effects of the relational operators (join $\bowtie$, projection $\pi$, and aggregate $\gamma$) on the data will also be \textit{shadowed} by the FFI calls. For instance, when \picachv is integrated with MySQL, its native query executors are invoked to process the corresponding node in the query plan. However, the native executor can only proceed to the next node if \picachv \textit{confirms} that its checks are passed, and no policy breaches are detected

\noindentsec{Sink.} After execution, a sink function (see \autoref{fig:query-execution}) is applied before results are returned. This step prevents data with remaining tags from inadvertently leaving the protected environment. This extra step is required to ensure operations must be performed on the data. Consider a scenario where an attacker attempts to access raw personal identifiers from a dataset that should first be aggregated. Permitting such data to pass through would breach the policy. Therefore, we implement a sink function that checks for the presence of any tags on the data. Thus, this final safeguard ensures that only properly processed and policy-compliant data is released from the system, with the protection of the query execution phase where \textit{disallowed operations} should never be performed.

\noindentsec{Verifiability.} The primary motivation for leveraging TEEs to host \picachv is to provide verifiable guarantees for private data processing on the server side to external parties (data owners) who lack direct control over their data in the cloud. TEEs offer data owners the ability to verify:

\begin{itemize}
\setlength\itemsep{0em}
    \item \textit{What} is being executed within the TEE by utilizing \textit{remote attestation}, which produces a non-forgeable cryptographic report detailing the TEE’s properties.
    \item \textit{How} the binary running inside the TEE was built, ensuring it originates from verified source code through a trusted build system.
\end{itemize}

By initiating a remote attestation session, data owners can validate the source code and build process referenced in the attestation report, providing reassurance that their data is being processed securely and as intended.

\subsection{Optimization Techniques}
\label{subsec:optimization}
The nature of \textsc{Picachv} as a dynamic security monitor naturally raises performance concerns due to the overhead of managing security labels at the cell level. We detail in this section potential optimization techniques.


\noindentsec{Materialization and caching.} Inspired by materialization techniques commonly used in databases~\cite{materialization, caching}, we adapt this concept for policy checking in \textsc{Picachv}. Our approach caches the verifications of frequently queried plans or sub-plans that have already been confirmed as compliant with data use policies. When a new query is issued, query rewriting techniques are applied to determine if the query can be transformed to leverage these cached materialized views. This strategy reduces redundant checks and significantly reuses prior verification computations, leading to improved efficiency. However, implementing this approach poses challenges for traditional program analysis and verification methods, as they typically lack the higher-level semantic abstractions necessary to capture the intent of programs.



\noindentsec{Hybrid scheme.} One of the key advantages of leveraging query plans is the inherent ability to gain deep insights into data flow. Query plans are embedded with strong semantics that detail not only the flow of data but also the types of operations performed, dependencies among data elements, and potential interferences. This rich semantic information allows us to conduct preliminary static analysis before engaging in dynamic verification processes. By performing static analysis on the query plan, we can potentially verify the entire query or its subcomponents before execution. If the static analysis successfully validates the entire query, then one can proceed to the next query without further checks. Moreover, in cases where the static analysis is insufficient or incomplete, dynamic verification can be seamlessly initiated from the point where the static analysis has left. 


\section{Evaluation}
\label{sec:evaluation}
In this section, we present the experimental results of \picachv to demonstrate that 1) The additional runtime overhead of \picachv is small, 2) \picachv can support many real-world analytical tasks and policies, and 3) \picachv can enforce these privacy policies.

\subsection{Experiment Setup}
\noindentsec{Test environment.} Our evaluation was conducted inside a VM TEE that is based on the Intel Trusted Domain eXtension (TDX) on a server with two 2.3 GHz Intel Xeon Platinum 8568Y+ CPUs (a total of 96 cores and 192 threads) and 512 GB of memory, running Ubuntu 22.04. We have integrated \picachv into a state-of-the-art data analytical engine called \texttt{Polars}~\cite{polars} (31K stars on GitHub): A powerful library for high-performance data manipulation and analysis in Python and Rust. We build all the components in release mode with optimization level at \texttt{O3}.

\noindentsec{Dataset and test suite.} 
In our benchmark, we employ the latest TPC-H specification (v3.0.1)~\cite{tpc-h}, using \texttt{tpch-dbgen}~\cite{tpch-dbgen} to generate a dataset of different sizes by changing the scale factors. The TPC-H is a decision support benchmark. It consists of a suite of 22 business-oriented \textit{ad hoc} queries on data split across 8 tables. The queries and the data populating the database have been chosen to have broad industry-wide relevance. We implement TPC-H queries in \texttt{Polars} (using its dataframe APIs). Currently, there are no official data use policies for the TPC-H testbed. To address this gap, we manually crafted policies to simulate real-world scenarios. We then manually verified output correctness.

\subsection{Performance Overhead}
\begin{figure}[h]
    \centering
    \begin{subfigure}[t]{0.23\textwidth}
        \centering
        \includegraphics[width=\textwidth]{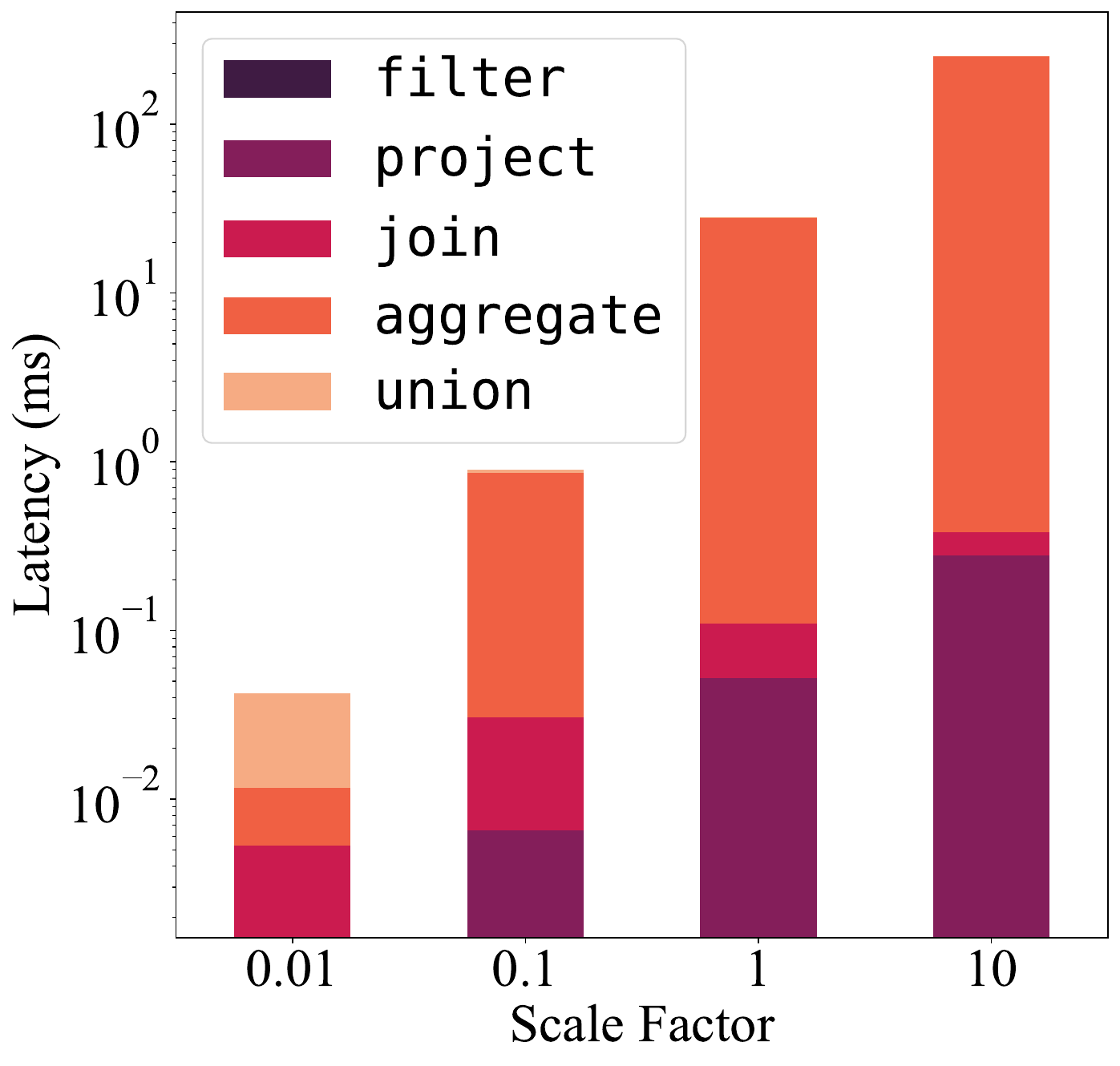}
        \caption{Policy \#A: Everything can be used.}
    \end{subfigure}%
    ~ 
    \begin{subfigure}[t]{0.23\textwidth}
        \centering
        \includegraphics[width=\textwidth]{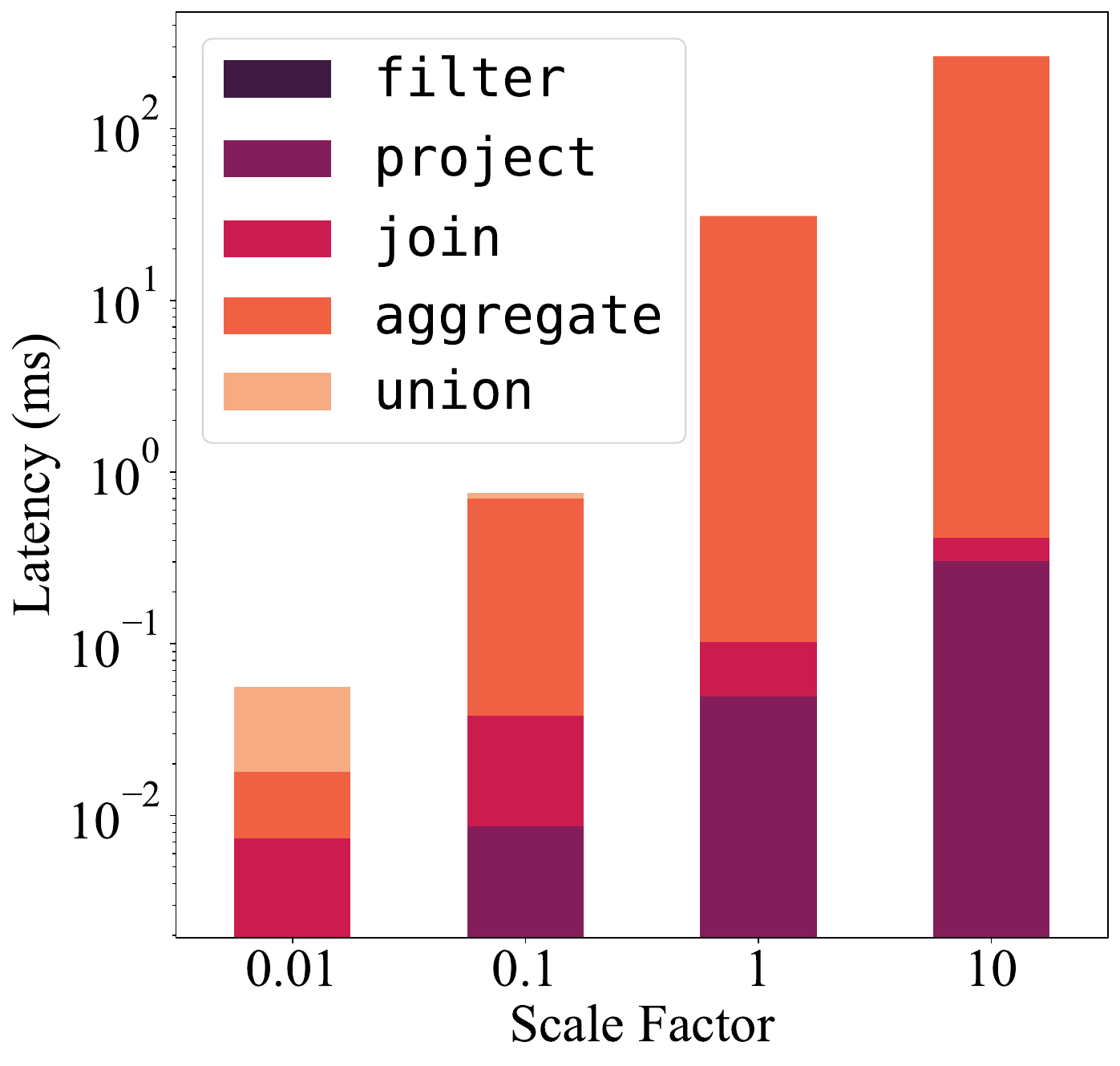}
        \caption{Policy \#B: The column \texttt{l\_discount} should be aggregated with group size > 5.}
    \end{subfigure}\\
    \begin{subfigure}[t]{0.23\textwidth}
        \centering
        \includegraphics[width=\textwidth]{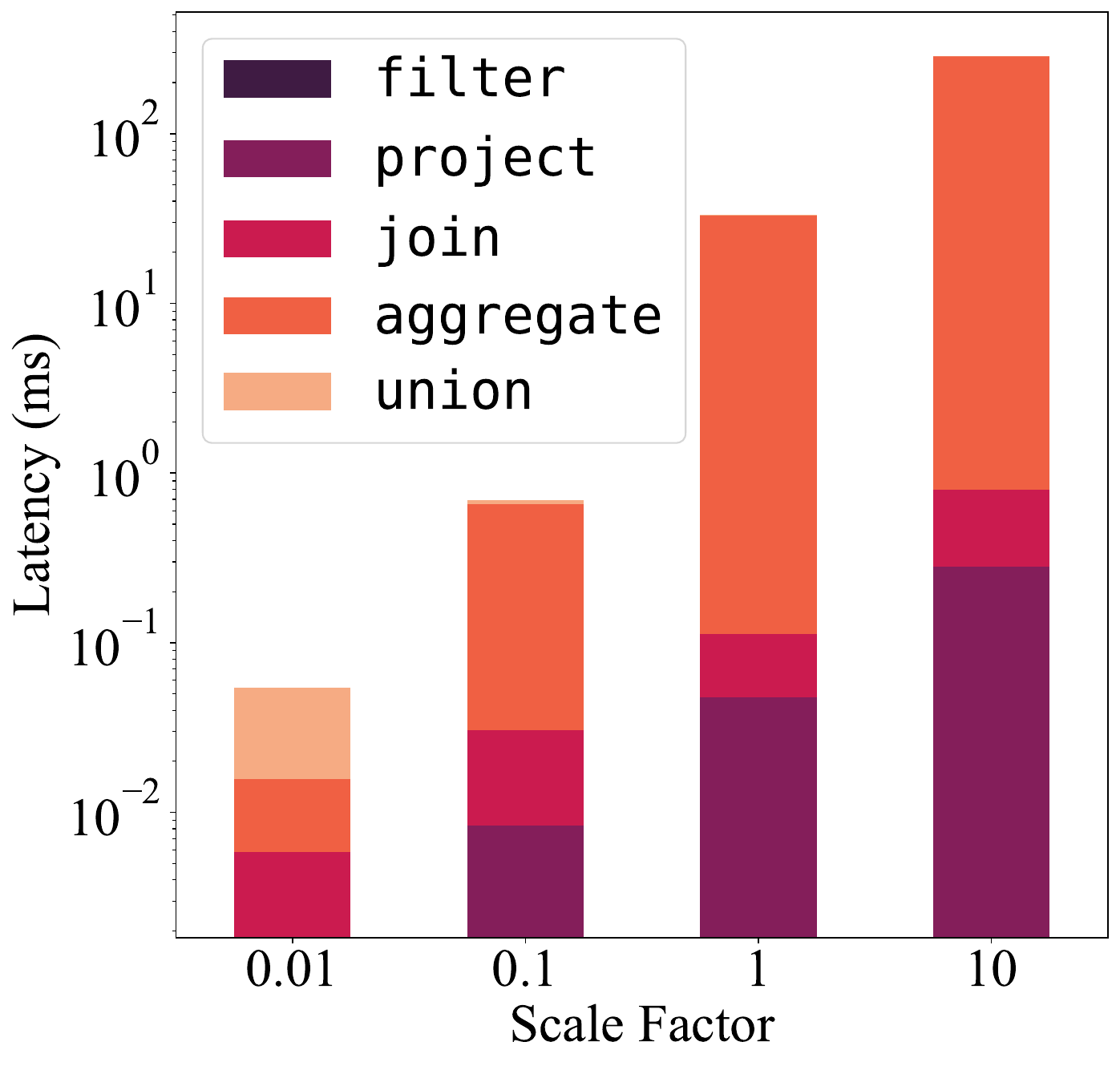}
        \caption{Policy \#C: The column \texttt{l\_discount} should be added by $1.0$.}
    \end{subfigure}
    ~
    \begin{subfigure}[t]{0.23\textwidth}
        \centering
        \includegraphics[width=\textwidth]{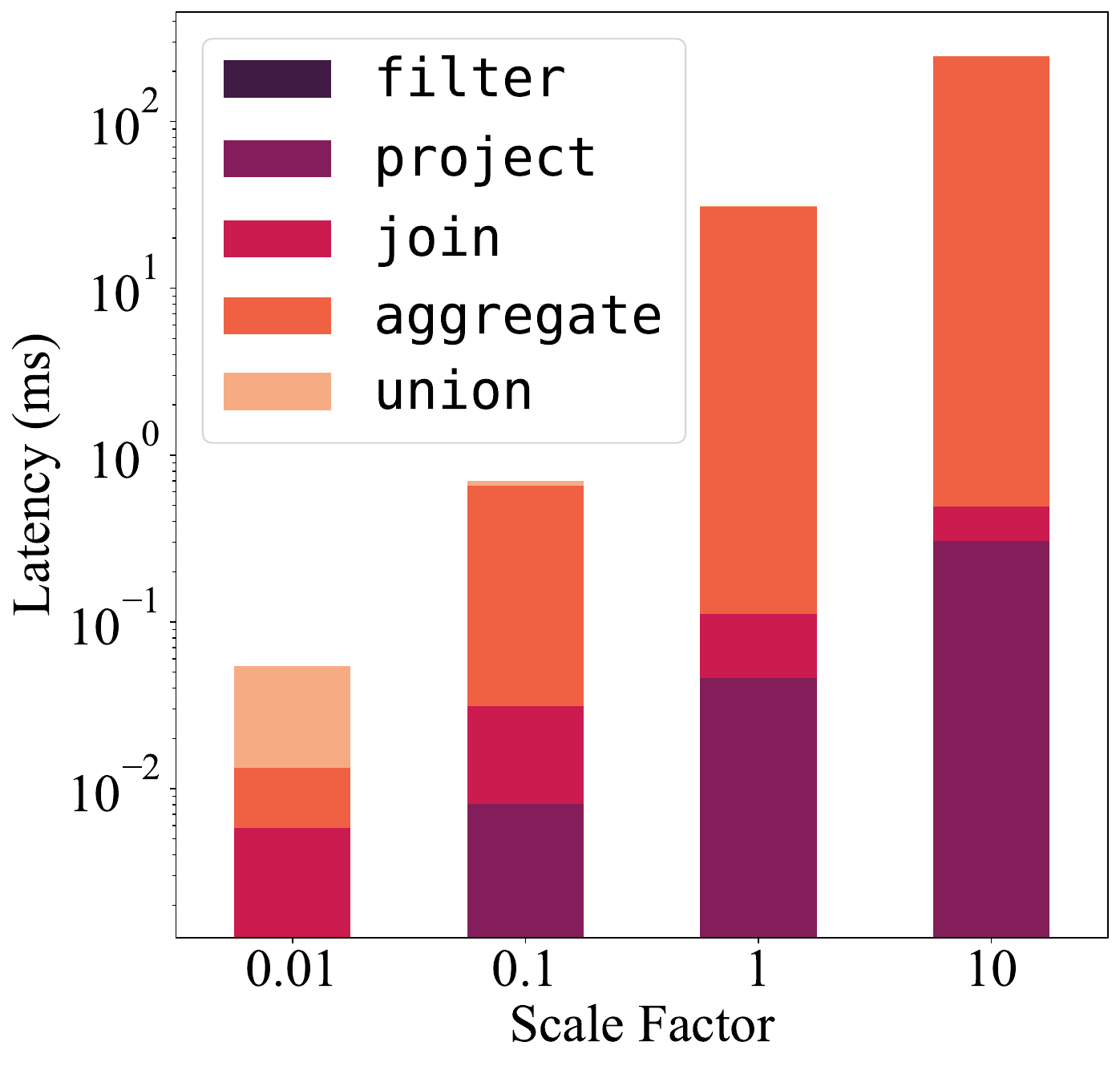}
        \caption{Policy \#D: Composition of Policy \#B and Policy \#C. }
    \end{subfigure}
    \caption{The result of the microbenchmark on each relational operator's runtime overhead.}
    \label{fig:microbenchmark}
\end{figure}

\begin{figure*}[]
    \centering
    \includegraphics[width=\textwidth]{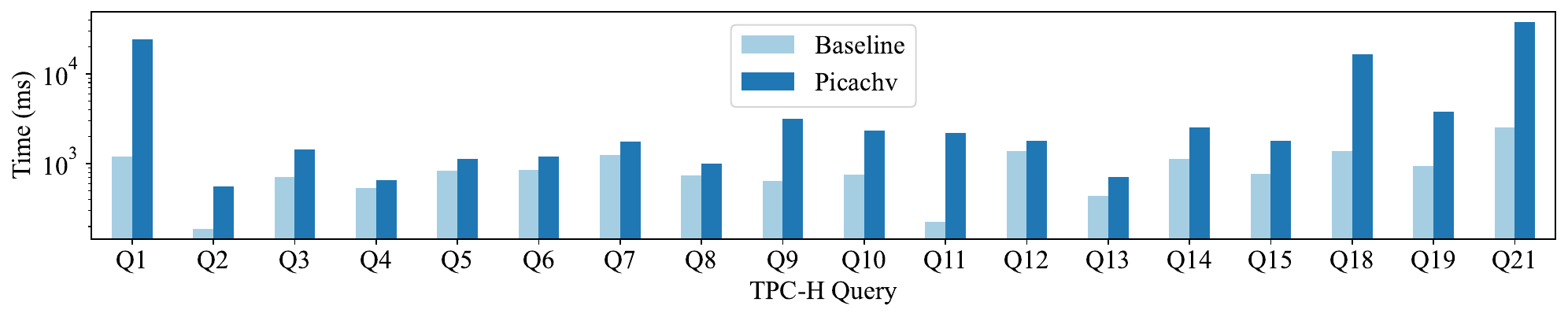}
    \caption{The runtime overhead of TPC-H testbed of \picachv.}
    \label{fig:macrobenchmark}
\end{figure*}

\noindentsec{End-to-end latency.} We begin by presenting an overview of \picachv's end-to-end performance overhead, comparing it against the unmodified query engines from \texttt{Polars} as the insecure baseline. We evaluate performance using selected queries from the TPC-H benchmark suite. In this experiment, we set the scale factor to 1. Queries not included in the benchmark contain some features currently not yet supported by \picachv. We exclude the policy file reading time from our measurements to provide a more accurate representation of runtime performance. We believe this overhead can be mitigated through strategies such as preloading policies during startup, and this often occurs infrequently. To measure the end-to-end performance of query execution with policy checking enabled, we assign dummy labels ($\bf L$) to each cell in this benchmark. \autoref{fig:macrobenchmark} shows the complete experimental results. \picachv generally shows higher execution times than the baseline (from $\sim 1.2 \times$ to $\sim 15 \times$), indicating some overhead from policy checking. Some queries, like Q8, Q12, and Q13, show minimal differences between Picachv and the baseline, while others, such as Q9, Q15, exhibit more noticeable performance gaps. Such large overheads, as indicated in the microbenchmark (see \autoref{fig:microbenchmark}), can be attributed to both the \texttt{projection} and \texttt{aggregate} operators.

\begin{table}[htbp]
    \centering
    \begin{tabular}{cc}
        \hline
        \bf Table size & \bf Policy file loading time (s) \\ \hline
         $10$ MB & 1.67 \\
         $100$ MB & 12.16 \\
         $1$ GB & 91.50 \\
         $10$ GB & 703.64 \\
         \hline
    \end{tabular}
    \caption{Time used to load policy files.}
    \label{tab:loading}
\end{table}

\begin{table*}[]
    \centering
    \begin{tabular}{lllll}
            \hline
            \textbf{Dataset used in the task} &\textbf{Task Name} & \textbf{Execution Time (ms)} & \textbf{Checking Time (ms)} & \textbf{Result} \\\hline
            \multirow{3}{*}{\makecell{Chronic illness~\cite{kaggleChronicIllness}}} & \multicolumn{1}{l}{1: autoimmune~\cite{kaggleAutoimmunesymptom}} & \multicolumn{1}{l}{$32.69$} & \multicolumn{1}{l}{$469.394\ (14.36 \times)$} & \multicolumn{1}{l}{\ding{55}; \texttt{user\_id}}
            \\\cline{2-5}
                                 & \multicolumn{1}{l}{2: EDA~\cite{kaggleFlaredownData}} & \multicolumn{1}{l}{$23.93$} & \multicolumn{1}{l}{$1416.04 (59.17 \times)$} & \multicolumn{1}{l}{\ding{55}; group size too small}\\\cline{2-5}
                                 & \multicolumn{1}{l}{3: Symptoms~\cite{kaggleFlaredownAutoimmune}} & \multicolumn{1}{l}{$170.25$}  & \multicolumn{1}{l}{$588.089 \ (3.46\times)$} & \multicolumn{1}{l}{\ding{51}}\\\hline
            \multirow{3}{*}{\makecell{Healthcare Dataset ~\cite{kaggleHealthcareDataset}}} & \multicolumn{1}{l}{4: Healthcare~\cite{kaggleAnalysisHealthcare}} & \multicolumn{1}{l}{$22.12$}
            & \multicolumn{1}{l}{$299.95\ (13.56 \times)$} & \multicolumn{1}{l}{\ding{51}}
            \\\cline{2-5}
                                 & \multicolumn{1}{l}{5: Trends~\cite{kaggleUnlockingHealthcare}} & \multicolumn{1}{l}{$49.58$} & \multicolumn{1}{l}{$132.55\ (2.67 \times)$}
                                 & \multicolumn{1}{l}{
                                 \ding{51}}
                                 \\\cline{2-5}
                                 & \multicolumn{1}{l}{6: Analysis~\cite{kaggleHealthCare}} & \multicolumn{1}{l}{$57.02$}& \multicolumn{1}{l}{$598.94\ (10.50\times)$}& \multicolumn{1}{l}{\ding{51}} \\\hline
        \end{tabular}
    \caption{Results of analytical tasks in the case study are presented. We use \ding{51} to indicate programs that comply with the privacy policy and \ding{55} to denote instances where a policy breach was detected. For programs that fail to meet policy requirements, we provide explanations. The original code was written in Python and subsequently adapted to Rust. We recorded execution times without policy enforcement to establish a baseline for comparison.}
    \label{tab:case-study}
\end{table*}

\noindentsec{Microbenchmark.} To understand what components contribute to the major runtime overhead, we choose query Q3\footnote{Since TPC-H does not explicitly include a \texttt{UNION} operation, we slightly modify Q3 to incorporate it.}  because the minimal query incorporates all the necessary operations we want to evaluate from the TPC-H benchmark, and we slightly modified it to accommodate the policy. We run it atop \texttt{Polars} to analyze the breakup of runtime overhead. The benchmark results are reported in \autoref{fig:microbenchmark}. To give a clearer understanding of which component significantly contributes to the overhead, we split the overhead into the following parts: 1) the cost of policy file loading, 2) \texttt{project}, 3) \texttt{join}, 4) \texttt{aggregate}, 5) \texttt{union}, and 6) \texttt{filter}. We also designed this query for different privacy policies to see which kinds of policies will cause significant policy checking overhead. Also note that in the microbenchmark, we do not consider the time taken for \texttt{polars} to fulfill the query but solely consider the time taken by \picachv.  In this experiment, we set the scale factor of the database generation from $0.01$ to $10$ (sizes from 10 MB to 10 GB) to test the performance under different sizes. In our experiment, we observed considerable overhead when loading the policy files from the disk, and we report the latency in \autoref{tab:loading} to give a clearer understanding of the overhead incurred by policy checking at runtime. Despite such a large overhead, we believe it can be reduced by utilizing more advanced optimization techniques. We instead focus on the runtime overhead of policy checking of \picachv. The experimental results in \autoref{fig:microbenchmark} show that the major performance overhead comes from the \texttt{project} and \texttt{aggregate} operator that involves data transformation, and as data grows, the percentage of the \texttt{aggregate} operator soon dominates (from $1\%$ to nearly $99\%$). Interestingly, the impact of privacy policy types is minimal during our evaluation.

\begin{figure}[h]
    \centering
    \begin{subfigure}{0.47\linewidth}
        \includegraphics[width=\textwidth]{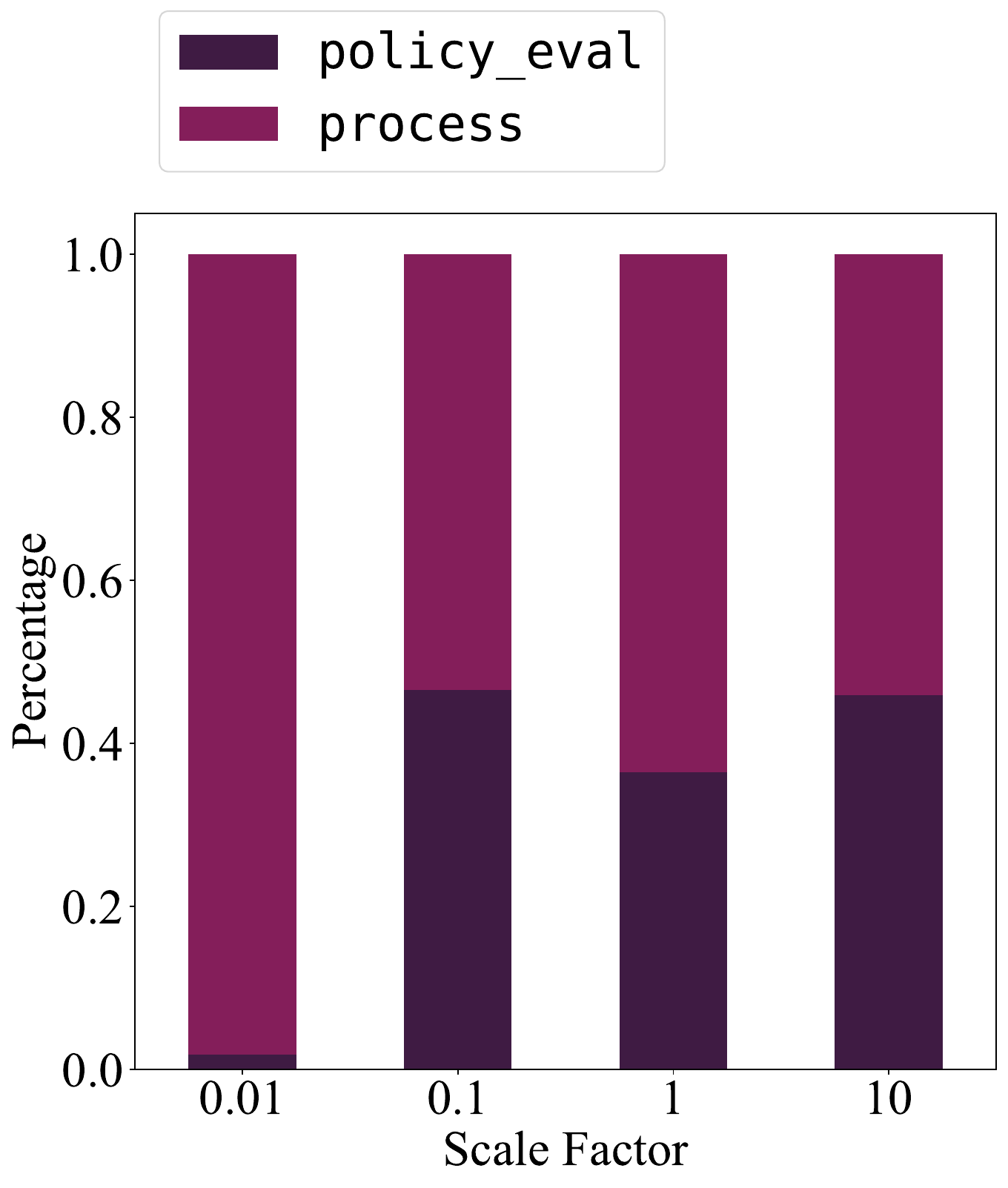}
        \caption{The cost breakdown of the \texttt{project} operator.}
        \label{fig:cost-proj}
    \end{subfigure}%
    \hfill
    \begin{subfigure}{0.47\linewidth}
        \includegraphics[width=\textwidth]{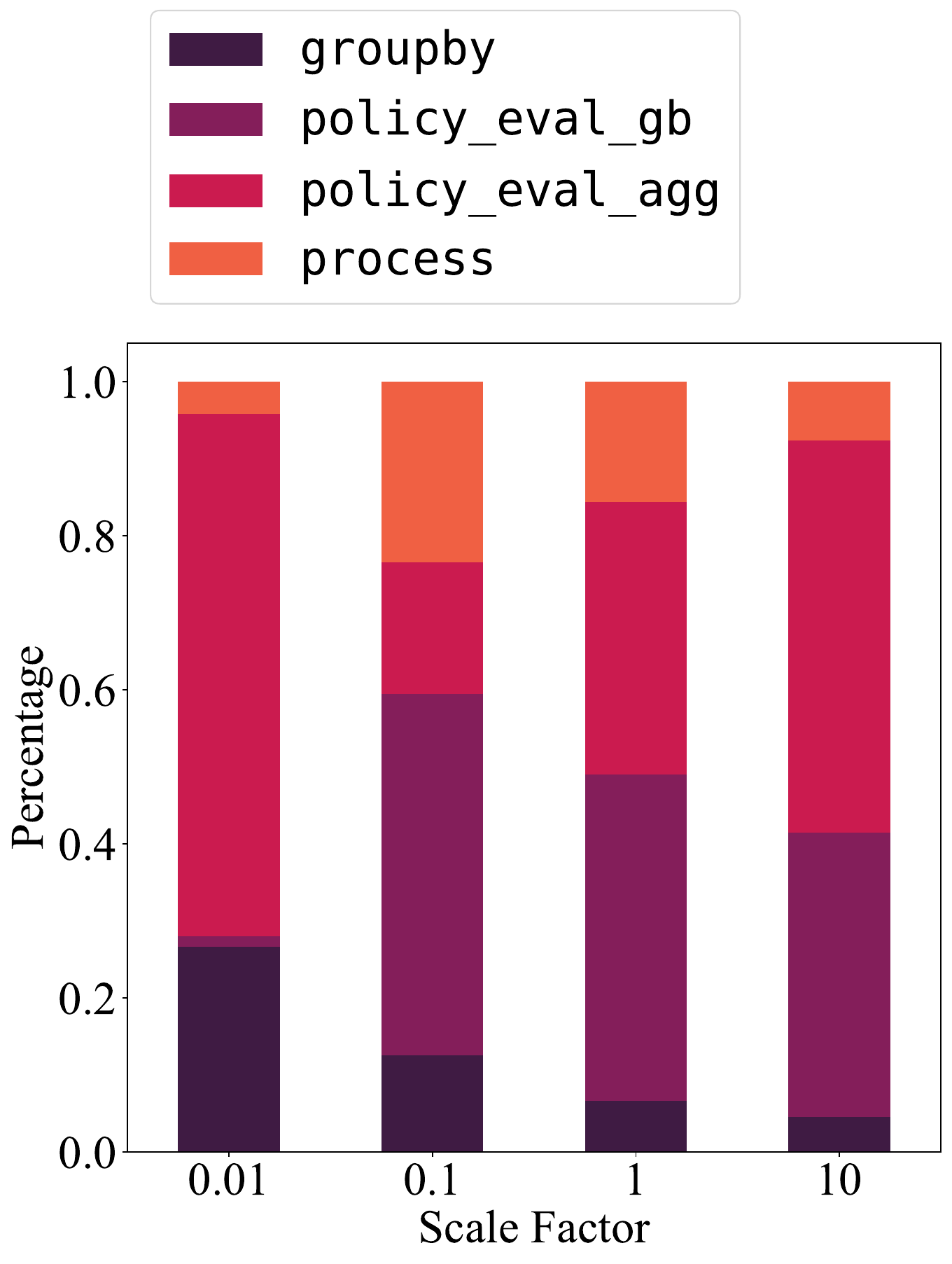}
        \caption{The cost breakdown of the \texttt{aggregate} operator.}
        \label{fig:cost-agg}
    \end{subfigure}
    \caption{Detailed analysis on the \texttt{project} and \texttt{aggregate} operators in terms of the percentage of their sub-routines.}
    \label{fig:proj-and-agg}
\end{figure}
\noindentsec{Analysis on \texttt{project} and \texttt{aggregate} operators.} To understand \textit{why} \texttt{project} and \texttt{aggregate} contribute significantly to runtime overhead, we conducted a detailed benchmark for these operators. In this experiment, we fixed the policy type of the composition of policy \#B and policy \#C to isolate their impact. In \picachv’s implementation, the \texttt{project} operator consists of two sub-routines: a) \texttt{policy\_eval}, which applies declassification rules to policy tags, and b) \texttt{process}, which handles intermediate in-memory representations for further processing. The \texttt{aggregate} operator includes: a) \texttt{group by}, grouping tuples as per the query plan; b) \texttt{policy\_eval\_gb}, applying declassification rules during grouping; c) \texttt{policy\_eval\_agg}, declassifying grouped policies; and d) \texttt{process}. Shadow tables are already in-memory, so disk I/O overhead is excluded. We decompose these operators into their primitives and provide a cost breakdown in \autoref{fig:proj-and-agg}. For the \texttt{project} operator (\autoref{fig:cost-proj}), \texttt{process} consistently dominates runtime overhead across all scale factors, while \texttt{policy\_eval} remains a smaller but noticeable contributor, especially at smaller scales. This shows the primary overhead in \texttt{project} comes from in-memory processing logic rather than policy evaluation. For the \texttt{aggregate} operator (\autoref{fig:cost-agg}), the experimental result highlights the increased complexity of its sub-routines. As the scale factor grows, the \texttt{process} ’s contribution remains prominent, but the costs of \texttt{policy\_eval\_gb} and \texttt{policy\_eval\_agg} also grow proportionally. Notably, \texttt{group by} contributes a smaller share to the overall latency because this operation can be efficiently parallelized. Since \texttt{aggregate} requires folding on the groups, This indicates that the policy evaluation stages in \texttt{aggregate} are computationally intensive and scale-sensitive. This explains why aggregate-intensive queries in the TPC-H testbed might have more runtime overhead. Future works can be focused on designing more efficient aggregate algorithms.

\subsection{Case Studies}
To demonstrate \picachv's practical applications and versatility, we conducted case studies using two datasets: a chronic illness dataset and a healthcare dataset. We adapted three analytical tasks for each dataset, implementing them using \texttt{Polars}. For these tasks, we manually crafted privacy policies simulating real-world scenarios such as those found in the All-of-Us project~\cite{allofus} and HIPAA regulations~\cite{hipaa}. The results, presented in \autoref{tab:case-study}, show that \picachv successfully detected policy violations in two out of six tasks, while allowing the compliant tasks to proceed. Notably, the policy checking time often exceeded the task execution time, and the in the worst case this would introduce nearly $59.17\times$ overhead, suggesting potential for future optimization. We found this is mainly due to the \texttt{group by} operation found commonly in data analytics, as suggested by \autoref{fig:microbenchmark} and \autoref{fig:cost-agg}.

These case studies validate \picachv's capability to enforce complex privacy policies particularly in sensitive domains like healthcare. The monitor's ability to detect and prevent policy violations, even for quickly executed tasks, underscores its value in ensuring responsible data use. Furthermore, the application of \picachv across diverse analytical tasks demonstrates its potential for adaptation to various data-intensive fields requiring stringent privacy protection.







\section{Discussion}
\label{sec:discussion}
\vspace{0.25em}\noindent\textbf{Semi-structured and unstructured data.} \textsc{Picachv}'s operation at the relational algebra level inherently limits its support for semi-structured and unstructured data. This limitation is significant, as many real-world workloads involve interactions with such data types. For instance, electronic health record (EHR) analysis often requires processing clinical notes, which are unstructured free text. Similarly, human genome analysis deals with semi-structured Single Nucleotide Polymorphism (SNP) data. Enforcing privacy policies on these diverse data formats presents a substantial challenge that extends beyond \textsc{Picachv}'s current capabilities. Addressing this limitation would greatly enhance the system's applicability across a broader range of data-intensive domains.

\vspace{0.25em}\noindent\textbf{Expressiveness of relational algebra.} While \textsc{Picachv}'s use of relational algebra as an abstraction enables policy enforcement for a wide range of data operations, it also presents limitations in expressiveness for certain advanced analytics tasks. Many modern data analytics workflows, particularly in machine learning and artificial intelligence, involve operations that extend beyond the capabilities of traditional relational algebra. For instance, complex matrix operations, iterative algorithms, and non-linear transformations common in ML models cannot be directly expressed using standard relational operators. Recent work has proposed primitive operators for a tensor relational algebra~\cite{tra} tailored to these specific use cases, analogous to SPJUA in traditional relational algebra. Future work could explore extending \textsc{Picachv}'s current solution to incorporate such algebras, potentially broadening its applicability to more advanced analytics scenarios.


\vspace{0.25em}\noindent\textbf{Complete verification.} We currently trust the query planner to produce correct and policy-compliant plans. However, this trust assumption introduces a potential vulnerability in the overall system. In reality, a comprehensive security guarantee would require verification of the entire pipeline, including the query planner. Verifying the planner would ensure that the generated query plans themselves adhere to the specified policies and do not introduce unexpected data flows or operations that could violate privacy constraints. This extension of our verification scope represents an important avenue for future work, as it would close a significant gap in the end-to-end formal guarantees of our system.

\vspace{0.25em}\noindent\textbf{Automated policy interpretation.} Translating privacy regulations like GDPR into computer-interpretable policies typically requires significant human effort, and this work implicitly assumes the existence of such policies. Recent advancements in Natural Language Processing (NLP), such as \textsc{ARC}~\cite{auto-regu}, offer promising solutions for automating this process. These technologies could bridge the gap between human-readable regulations and machine-executable policies, enhancing efficiency and accuracy in applying privacy standards. While crucial for privacy compliance, we consider this challenge orthogonal yet complementary to \textsc{Picachv}. The synergy between automated policy interpretation and \textsc{Picachv} could significantly advance privacy-preserving data analytics.
\section{Conclusion}
\label{sec:conclusion}

In this paper, we present \textsc{Picachv}, a significant advancement in enforcing data use policies for analytics. By abstracting program semantics via relational algebra and employing formal verification, we have created a system that effectively balances policy compliance with analytical flexibility. Our evaluations demonstrate \textsc{Picachv}'s efficiency, accuracy, and real-world applicability across various regulatory frameworks. While limitations exist, particularly for non-relational data, \textsc{Picachv} provides a robust foundation for responsible data usage in an increasingly data-driven world. This work paves the way for future developments in secure data analytics.

\section*{Acknowledgements}
The authors would like to thank our shepherd and other anonymous reviewers for their constructive feedback. The authors would also like to thank Prof. Danfeng Zhang for the fruitful discussion on the topic of information flow control, members of CDCC for their comments on the early draft of this paper, and Haosen Guan for his support. This work was supported in part by National Science Foundation (NSF) Grant No. 2207231. Any opinions, findings, conclusions, or recommendations contained in this material are those of the authors and do not necessarily reflect the views of the National Science Foundation.
\section*{Ethics Considerations}
The authors of this paper carefully reviewed related documents and \textsc{Picachv} in its design and implementation. As a policy enforcement tool, the design and implementation do not involve any ethical considerations.

\section*{Open Science}
All the data and programs used in the evaluation section are publicly available. Furthermore, the authors will fully disclose the source code, formal proofs, and other benchmark tools as a standalone artifact to the public to support future research. Code can be found at \url{https://github.com/picachv}.

\bibliographystyle{plain}
{\footnotesize\bibliography{ref}}
\appendix
\section{Proofs}
\subsection{Proofs for \autoref{thm:soundness}}
\begin{proofsketch}
    Let $\Sigma$ be any data store, $q$ be any query, and $st$ be any program state. By definition, either $\Sigma \vdash q \ \textcolor{blue}{\Downarrow}\ \tuple{R, tr}$ holds or it does not. For the former case, we apply mathematical induction over $\Sigma \vdash q \ \textcolor{blue}{\Downarrow}\ \tuple{R, tr}$. In the latter case, an error is thrown and nothing is returned.
\end{proofsketch}
\subsection{Proofs for \autoref{thm:non-inter}}
\begin{proof}
    By the definition of the finalization function at the end of the execution, we immediately filter out data with the remaining tags. Thus, if the program returns valid data, then following the soundness theorem, due to invocation of the sink function, we know that $\forall c \in R, \mathcal{E}(c) \equiv \mathbf{L}$, meaning that the query trace is equivalent to a computation that involves no secret. This completes the proof.
\end{proof}
\begin{table*}[t]
    \centering
    \begin{tabular}{cc}
         \hline
         \bf Dataset & \bf Privacy Policy in Natural Language\\
         \hline
         Chronic illness~\cite{kaggleChronicIllness} & \makecell{1. People whose \texttt{age} > 89 must be generalized (Safe Harbor). \\ 2. \texttt{user\_id} should be removed (Safe Harbor). \\ 3. \texttt{trackable\_*} should only be aggregated with one of \texttt{MAX, MIN, SUM, COUNT}, and \\ size should be greater than 20 (NIH-like policy).}   \\\hline
         Healthcare dataset~\cite{kaggleHealthcareDataset} & \makecell{1. \texttt{name} should be removed (Safe Harbor). \\ 2. \texttt{medical\_condition} must be aggregated with one of \texttt{MAX, MIN, SUM, COUNT}. \\ (Common aggregate requirements)} \\\hline
    \end{tabular}
    \caption{The dataset and its corresponding privacy policies used in case studies.}
    \label{tab:dataset-and-policy}
\end{table*}
\section{Policies Used in the Case Studies}
The policies used in the case study of \autoref{sec:evaluation} are presented in \autoref{tab:dataset-and-policy}.


\section{Query 3 from the TPC-H Benchmark}
We present the code of Query 3 from the TPH-H benchmark used in our microbenchmark in \autoref{lst:query3}.

\begin{lstlisting}[label=lst:query3, language=sql, caption=Code of Query 3 in the TPC-H Benchmark, basicstyle=\ttfamily\small]
SELECT
    l_orderkey,
    sum(l_extendedprice * (1 - l_discount)) as revenue,
    o_orderdate,
    o_shippriority
FROM
    customer,
    orders,
    -- We added a self-union for `lineitem`.
    (SELECT * FROM lineitem)
        UNION ALL
    (SELECT * FROM lineitem)
WHERE
    c_mktsegment = 'BUILDING'
    AND c_custkey = o_custkey
    AND l_orderkey = o_orderkey
    AND o_orderdate < date '1995-03-15'
    AND l_shipdate > date '1995-03-15'
GROUP BY
    l_orderkey,
    o_orderdate,
    o_shippriority
ORDER BY
    revenue desc,
    o_orderdate
LIMIT 20;
\end{lstlisting}

\end{document}